\documentclass[journal]{IEEEtran}

\usepackage{caption}
\usepackage{amsmath}
\usepackage{cite}
\usepackage{pifont}
\usepackage{url}  
\usepackage[pdftex]{graphicx}
\usepackage{hyperref}
\usepackage[inline]{enumitem}
\usepackage{graphicx}
\usepackage{multirow}
\usepackage{amssymb}
\usepackage{threeparttable}
\usepackage{booktabs}
\usepackage{subcaption}
\usepackage{color}
\usepackage{textcomp}
\usepackage{amsthm}

\newtheorem{assumption}{Assumption}
\newtheorem*{remark}{Remark}

\hypersetup{
	colorlinks = true, 
	urlcolor  = cyan, 
	linkcolor = blue, 
	citecolor = red 
}

\newcommand{\CS}{\text{CS}}
\newcommand{\EM}{\mathbf{EM}}
\newcommand{\dis}{\text{dis}}
\newcommand{\charge}{\text{char}}
\newcommand{\EV}{\text{EV}}
\newcommand{\arrive}{\text{arrive}}
\newcommand{\leave}{\text{leave}}
\newcommand{\unplug}{\text{unplug}}
\newcommand{\init}{\text{init}}
\newcommand{\amb}{\text{amb}}
\newcommand{\req}{\text{req}}
\newcommand{\sig}{\text{sig}}
\newcommand{\wh}{\text{wh}}
\newcommand{\ON}{\text{on}}
\newcommand{\OFF}{\text{off}}
\newcommand{\BSS}{\text{BSS}}
\newcommand{\temp}{\text{temp}}
\newcommand{\upd}{\text{upd}}
\newcommand{\all}{\text{all}}
\newcommand{\ds}{\text{ds}}
\newcommand{\real}{\text{real}}
\newcommand{\ES}{\text{ES}}
\newcommand{\SL}{\text{SL}}
\newcommand{\PV}{\text{PV}}
\newcommand{\ir}{\text{ir}}
\newcommand{\CST}{\text{CST}}
\newcommand{\APR}{\text{apr}}
\newcommand{\REV}{\text{rev}}
\newcommand{\LC}{\text{LC}}
\newcommand{\tE}{t=T}
\newcommand{\tS}{t=1}

\begin{document}

\title{Optimal Energy Dispatch of Grid-Connected Electric Vehicle Considering Lithium Battery Electrochemical Model}

\author{Yuanbo~Chen,~\IEEEmembership{Student Member,~IEEE,}
	Kedi~Zheng,~\IEEEmembership{Member,~IEEE,}
	Yuxuan~Gu,
	Jianxiao~Wang,~\IEEEmembership{Member,~IEEE,}
	and~Qixin~Chen,~\IEEEmembership{Senior Member,~IEEE}
\thanks{Manuscript submitted Dec. 15, 2022; revised May. 20 and Jul. 26, 2023; and accepted Sep. 16, 2023. This work was supported by the International (NSFC-NWO) Joint Research Project of National Natural Science Foundation of China under Grant 52161135201 and the Major Smart Grid Joint Project of the National Natural Science Foundation of China and State Grid under Grant U2066205.  \textit{(Corresponding author: Qixin Chen)}}
\thanks{Y.~Chen, K.~Zheng, Y.~Gu, and Q.~Chen are with the State Key Laboratory of Power Systems, Department of Electrical Engineering, Tsinghua University, 100084, Beijing, China. (e-mail:qxchen@tsinghua.edu.cn)}
\thanks{J.~Wang is with the National Engineering Laboratory for Big Data Analysis and Applications, Peking University, 100871, Beijing, China.}
\thanks{Digital Object Identifier \href{https://doi.org/10.1109/TSG.2023.3317590}{10.1109/TSG.2023.3317590}}
}

\markboth{IEEE TRANSACTIONS ON SMART GRID, VOL. XX, NO. XX, XXXX}
{Yuanbo~Chen \MakeLowercase{\textit{et al.}}: Optimal Energy Dispatch of Grid-Connected Electric Vehicle Considering Lithium Battery Electrochemical Model}
\maketitle

\IEEEpubidadjcol

\IEEEpubid{\begin{minipage}{\textwidth}\ \\[12pt] \centering
		© 2024 IEEE.  Personal use of this material is permitted.  Permission from IEEE must be obtained for all other uses, in any current or future media, including reprinting/republishing this material for advertising or promotional purposes, creating new collective works, for resale or redistribution to servers or lists, or reuse of any copyrighted component of this work in other works.
\end{minipage}}

\begin{abstract}
	The grid-connected electric vehicles (EVs) serve as a promising regulating resource in the distribution grid with Vehicle-to-Grid (V2G) facilities. In the day-ahead stage, electric vehicle batteries (EVBs) need to be precisely dispatched and controlled to ensure high efficiency and prevent degradation. This article focuses on considering a refined battery model, i.e. the electrochemical model (EM), in the optimal dispatch of the local energy system with high penetration of EVs which replenish energy through V2G-equipped charge station and battery swapping station (BSS). In this paper, to utilize the EM efficiently, recursive EVB constraints and a corresponding matrix-based state update method are proposed based on EM power characterization. The charging EV state distribution is profiled and a multi-layer BSS model along with binary aggregation is proposed, in order to overcome the computation complexity of combining the refined battery constraints with the mixed integer optimization. Finally, a local energy system scenario is investigated for evaluation. The efficiency and effectiveness of EM consideration are assessed from the perspective of both the system and battery.
\end{abstract}


\IEEEpeerreviewmaketitle

\section*{Acronyms}
\IEEEpubidadjcol

\begin{IEEEdescription}
	\item[BSS] Battery Swapping Station
	\item[CS] Charge Station
	\item[ECM] Equivalent Circuit Model
	\item[EM] Electrochemical Model
	\item[ES] Energy Storage
	\item[EV] Electric Vehicles
	\item[EVB] Electric Vehicle Battery
	\item[LiB] Lithium Battery
	\item[LES] Local Energy System
	\item[LFP] Lithiated Ferrum Phosphate
	\item[MILP] Mixed Integer Linear Programming
	\item[NCM] Nickel Cobalt Manganese
	\item[PV] Photovoltaic
	\item[SOC] State of Charge
	\item[SOPT] State of Power-Thermal
	\item[SSM] Source and Sink Model
	\item[V2G] Vehicle-to-Grid
\end{IEEEdescription}
	
\section{Introduction}
\IEEEpubidadjcol

The advancements in LiB, including the material innovation for higher capacity \cite{xie_retrospective_2020} and manufacturing improvements for cost reduction \cite{masias_opportunities_2021}, have aroused the popularization of EVs. As one of the key approaches to civil transportation electrification, EVs, however, have been cast doubt on the problem of ``range anxiety"\cite{coffman_electric_2017}. To combat such anxiety, besides the trend of higher LiB density, fast charging and battery swapping are the two coexisting approaches of fast energy replenishment as the alternative solution in different real-world applications \cite{liu_electric_2012}. 

In the meantime, with the progressing penetration of EVs in neighborhoods, uncoordinated charging loads have imposed heavy uncertainty on the distribution grid as well as the energy dispatch. Thus, serving as a coordinated interconnection between EVs and the grid, V2G technology has wide application in the distribution grid of high EV penetration scenario \cite{trivino-cabrera_joint_2019}. It can provide adjustable charging policy to modify the load curve\cite{zhang_evaluation_2017} and the potential of assimilating excess PV \cite{zheng_integrating_2019}.

Many researchers investigate the V2G potential based on CSs, among which V2G can be classified into a uni-directional and a bi-directional category. The uni-directional V2G only supports controllable charge other than discharge, enabling EVs to participate in the energy market as an adjustable load without extra equipment retrofit \cite{sortomme_optimal_2011}.

Whereas the bi-directional V2G considers the EVB as the automotive energy storage, providing regulation potential for the system operator\cite{zheng_integrating_2019}. The feasibility of penetrating a single V2G-equipped EV into a residential building is verified in \cite{monteiro_operation_2016}. While the large-scale EV fleet at V2G-equipped CS seems more attractive from grid prospects. \cite{zhang_evaluation_2017} estimates the V2G potential of the EV fleet by developing an aggregated model and a smart charging strategy. \cite{tabatabaee_stochastic_2017, chandra_mouli_integrated_2019} investigate the combination of bi-directional V2G and renewables to obtain the benefits of renewable assimilation as well as reduced cost.

Multiple grid-side ancillary services can find benefits with V2G grid-connected EVs, including demand response \cite{sortomme_optimal_2012, erdinc_smart_2015}, spinning reserve \cite{sortomme_optimal_2012}, and stability regulation \cite{liu_vehicle--grid_2015, kaur_coordinated_2019, hu_distributed_2022}. To address the issue of frequency stability, \cite{kaur_coordinated_2019} proposes a control strategy that utilizes the potential of V2G operations to participate in secondary frequency regulation. While in \cite{hu_distributed_2022}, voltage stability in the distribution network is investigated by employing reactive power V2G facilities.

Meanwhile, the commercialization of battery swapping in recent years has exhibited a more appealing V2G potential due to the fundamental LiB repository on site, which leads to abundant discussions on BSS scheduling. It is reported that V2G-equipped BSS has potential applications in scenarios such as peak-shaving\cite{esmaeili_optimal_2019}, frequency regulation\cite{wang_vehicle_2021}, and resilience provision\cite{najafi_efficient_2021}.

To instruct the LiB energy replenishment in BSS, \cite{wu_optimization_2018} studies the optimal operation of BSS based on standard LiB charging strategy and \cite{kang_centralized_2016} takes the priority and location of EV charging into account to propose a centralized charge policy. To deal with the computational complexity from the additional binary decision variables which are generated to illustrate the management of swapping LiB dispatch, \cite{you_optimal_2016} decouples the overall optimal charging decision into multiple independent subproblems by standard dual decomposition to implement parallel computing. To model the repository variation, \cite{liang_configuration_2021} develops a configuration of BSS LiB inventory to manage the swapping process and estimates the economic benefits, and \cite{ding_integrated_2021} provides an inventory-oriented BSS model to obtain the time-varying swapping price for the allocation. In \cite{ni_inventory_2021}, the charging center is separated from multiple distributed BSSs, and day-ahead inventory planning is proposed to determine the number of full-charged LiB delivery and depleted LiB recovery at the beginning and end of a day, respectively.

Regarding the V2G-equipped BSS, a few researchers have studied the power management of BSS from the aspect of a local grid, where the responsibility of selecting the economical spot of BSS is assigned to the system operator. \cite{sarker_optimal_2015} investigates the optimal behavior of the BSS under real-time pricing by formulating the internal operation as a MILP problem. In \cite{deng_hierarchical_2022}, the integration of on-site PV along with EV and BSS is studied on the day-ahead stage according to forecasted PV data. While in \cite{li_optimal_2018, esmaeili_optimal_2019}, the dispatch is divided into a double-layer framework to separately make decisions for microgrid and BSS with exchanged indicators of price and power.

Despite the advantages that BSS can bring to the grid, the main obstacle to promoting battery swapping is ultimately a commercial issue: a large amount of initial investment is required, and the contradictions of battery ownership between EV owners and BSS operator exist~\cite{ahmad_battery_2020}. Besides, EV manufacturers can hardly comprise to employ standardized battery pack structure design for swapping~\cite{zheng_electric_2014}. Therefore, it is estimated that among the energy replenishing modes, battery swapping will coexist with fast charging and mainly be engaged in to-business scenarios with the desire for replenishing energy rapidly for a relatively large capacity battery or a frequent demand, e.g., logistics trucks\cite{deng_hierarchical_2022}, public transportation \cite{you_optimal_2016} and autonomous mobility on demand (or taxis)\cite{ding_integrated_2021}. While fast charging will be provided for to-customer scenarios due to its universal application. 

From this perspective, instead of only considering the operation of BSSs alone like ~\cite{sarker_optimal_2015, esmaeili_optimal_2019,wu_optimization_2018,kang_centralized_2016,you_optimal_2016,liang_configuration_2021,ding_integrated_2021,deng_hierarchical_2022,li_optimal_2018}, the overall economic benefit of a microgrid consisting of both CS and BSS is investigated in \cite{zhong_cooperative_2022} through the Nash bargaining approach. \cite{liu_distributed_2019} focused on the logistic system between CS and BSS and solves the dispatch problem in a distributed way. Integrating CS and BSS, a game theoretic optimization framework is employed in \cite{ahmad_cost-efficient_2019} to coordinate the EVs and the optimal energy management considering the market. Furthermore, the business model of combined V2G-equipped CS and BSS along with the renewables and energy storage in the local grid has been also advocated by several leading EV infrastructure operators such as NIO \cite{noauthor_nio_2022}.

However, among the studies concerning CS and BSS, most of the existing literature focuses on the formulation of macro control logic neglecting the precise LiB model by setting a SOC level and assuming a constant charging period~\cite{kang_centralized_2016,liang_configuration_2021,hu_distributed_2022,ni_inventory_2021}, or with a simplified consideration on LiB model~\cite{tabatabaee_stochastic_2017, chandra_mouli_integrated_2019,sortomme_optimal_2012,erdinc_smart_2015,liu_vehicle--grid_2015,kaur_coordinated_2019, sarker_optimal_2015, esmaeili_optimal_2019,wu_optimization_2018,you_optimal_2016,ding_integrated_2021,deng_hierarchical_2022,li_optimal_2018,wang_vehicle_2021,najafi_efficient_2021}. The constant power limitation and direct update of SOC from power, i.e. SSM, is commonly applied, which leverages the LiB characteristic error from the actuality. A few related research on energy storage propose the application of the ECM~\cite{sakti_enhanced_2017}, but its parameters play a crucial role and need artificially modified under the varying operating situation.

Whereas in V2G-equipped CS and BSS, due to more frequent EVB charging and discharging, it is demanding for either SSM or ECM to capture precisely the variation of EVB physical characteristics, resulting in reduced battery efficiency, accelerated degradation, and even safety hazards. Thus there exists the necessity for a more refined LiB model. Fortunately, other than SSM and ECM, the EM of LiB originates from the description of the inside chemical reaction and excels in precise simulating results. By applying EM, some researchers managed to account for the degradation in optimal battery control \cite{reniers_improving_2018,cao_multiscale_2020} and to obtain a more authentic evaluation of battery power performance \cite{zheng_lithium-ion_2018,li_unlocking_2022}. Meanwhile, this mechanism-based model can reflect the internal characteristics to the dispatch level \cite{gu2022sop}, providing adaptability under different conditions and exploiting the capability of LiBs under the premise of safety.

But to the best of our knowledge, there has been no prior research investigating the optimal energy dispatch model of grid-connected EVs considering the EM, due to the sophisticated mechanism and the insufficiency of existing commercial solvers for such integrated complex problems. To enable this research, it is necessary to take reformulations and simplifications on both the EM and dispatch, which is key to handling the complexity of considering the complicated EM states in the solution of MILP which contains many logic integers.

Thus, motivated by the aforementioned research gap, this paper proposes an optimal dispatch model of grid-connected EVs at V2G-equipped CS and BSS, taking EM's advantage in interpreting battery performance and precise control. The main contributions of this paper are presented as follows:

\begin{enumerate*}[itemjoin=\\\hspace*{\parindent}]
	\item This paper accesses EM in energy dispatch through LiB power characterization to consider dynamic voltage, dynamic cell temperature, and dynamic available power. A matrix-based state update is proposed to avoid iterative calculation in the recursive constraints of EVB states.
	\item The EM power characteristics are incorporated efficiently with the grid-connected EVs at CS and BSS. Charging EVs' temporal behaviors are grouped and profiled to capture the uncertainty of CS state distribution without extra integers. While a multi-layer BSS model along with binary aggregation is presented to overcome the complexity of combining the recursive constraints of EVBs and binary control logic of BSS. 
	\item The case scenario is set in an LES with high EV penetration consisting of CS and BSS, exploiting the V2G potential of grid-connected EVs while effectively preserving high EVB efficiency and curtailing degradation.
\end{enumerate*}

The remainder of this article elaborates as follows: Section~\ref{sec:LPC} presents the methodology of obtaining power characteristics, whereas charging EV and BSS with EM are modeled in Section~\ref{sec:DispatchModel}. Section~\ref{sec:LES} establishes the optimal dispatch problem and Section~\ref{sec:case-study} carries out the case study. Finally, Section~\ref{sec:conclusion} delivers the conclusion of this article.

\section{Accessing the Power Characteristics of EM}
\label{sec:LPC}
\subsection{The Basic Idea}
\label{sec:PCTheory}
Denote the EM model from \cite{gu2021simplified} by $\EM(\cdot)$. The input current series are denoted by $ \boldsymbol{I} $ and material-related parameters by $ \boldsymbol{\psi} $. $ \boldsymbol{V} $ and $ \boldsymbol{\Theta} $ are sequential external states, i.e. voltage and cell temperature respectively. $ \boldsymbol{\Gamma} $ denotes the other internal electrochemical state matrices including lithium concentration $ \boldsymbol{\xi}_{s}^{\pm} $, cell energy conversion efficiency $ \boldsymbol{\eta} $, solution-solid interface potential $ \boldsymbol{\phi}_{se}^{-} $, active lithium loss $ \boldsymbol{\tau}_{L} $ and etc. $ SOC_0 $ and $ \Theta_0 $ are the initialization of SOC and cell temperature.

A straightforward idea for considering EM in dispatch is a two-level hierarchical model (M0) in (\ref{eqa:M0-Objective}), where the upper level solves the optimization and the lower level simulates the EM \cite{reniers_improving_2018}. However, optimizing M0 is impractical in this dispatch due to the computational burden. The state equations of EM can hardly be either incorporate with the iteration of MILP solving process or rewritten with analytic methods like Lagrange multipliers considering the following EM complexity:
\begin{enumerate*}[label=\itshape\alph*\upshape)]
\item high sampling frequency (\textasciitilde 1s);
\item non-linear and non-convex;
\item sequentially coupled.
\end{enumerate*}
\begin{equation}
\label{eqa:M0-Objective}
\begin{aligned}
	&\text{(M0)~minimize:}   \quad \text{Cost}\\
	&\begin{aligned}
		\text{subject to:} &~\text{Constraints~(EVB~operating~conditions,}\\
		&~\text{EV~distribution,~BSS~logic,~etc)} \\
		&~(\boldsymbol{V},~\boldsymbol{\Theta},~\boldsymbol{\Gamma})= \EM(SOC_0, \Theta_0, \boldsymbol{I}, \boldsymbol{\psi})
	\end{aligned}
\end{aligned}
\end{equation}

Therefore, the EM needs further simplifications through LPC to be considered in dispatch. The core idea of the LPC is to extract generic power-related state relationships and generic boundary conditions between adjacent decision points in the optimization. It is essentially a mapping of EM on the low-dimension space of the dispatch, i.e. dimensional reduction.

\begin{figure}[htb]
	\centering
	\includegraphics[width=1\linewidth]{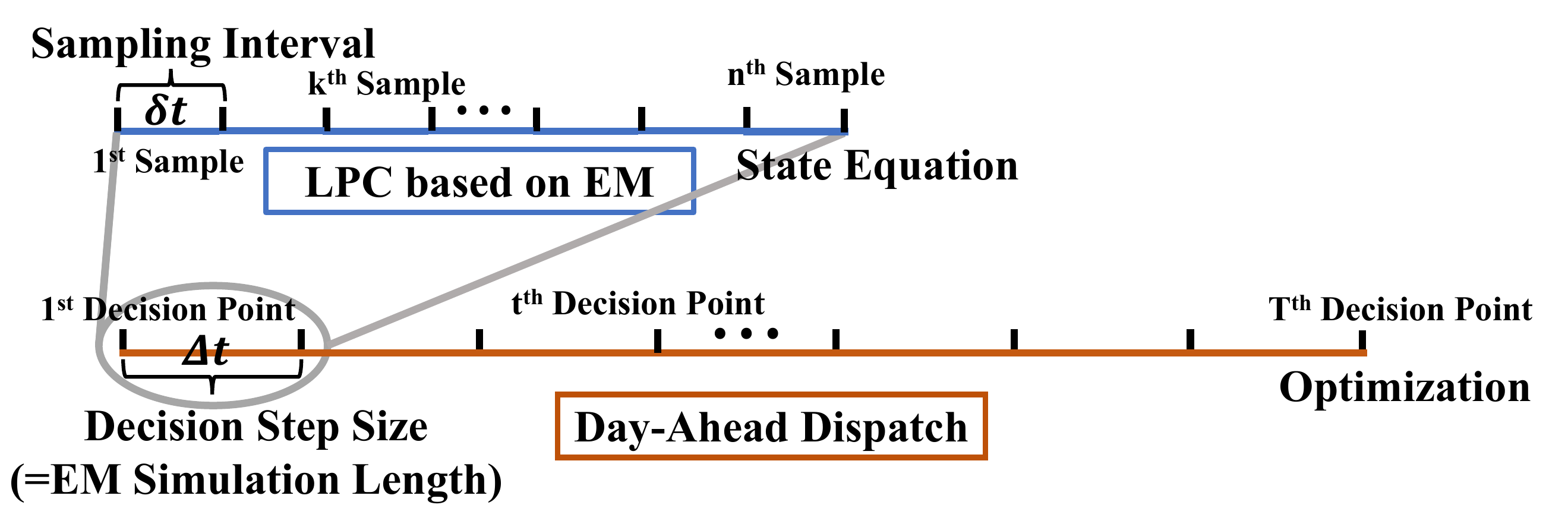}
	\caption{Time Resolution of LPC and Day-Ahead Dispatch}
	\label{fig:power_characterization}
\end{figure}

As shown in Fig.~\ref{fig:power_characterization}, LPC is performed on high-sampling frequency EM (second-level) to retain EM accuracy in terms of interpreting battery performance. The scope of LPC should be equal to the decision step size, to eliminate the varying influence of LPC time duration on battery performance. Besides, the selection of decision step length actually balances the computational complexity between the LPC and the energy dispatch problem, i.e., a shorter decision step length decreases the EM simulating time in LPC but increases the number of decision variables in optimization.

This article addresses the concerning LiB performance in energy dispatch, including the dynamic voltage, dynamic cell temperature, and dynamic available power of EVB, which are implemented by power dynamics, heat dynamics, and SOPT, respectively. Drawing inspiration from the numerical method presented in \cite{gu2022sop}, this article enhances the LPC framework and proposes the recursive constraints and matrix-based state update to efficiently integrate LPC with MILP optimization.

\subsection{LPC Framework}
\subsubsection{Power dynamics}
\label{sec:Power dynamics}
Power dynamics aim to consider the dynamic voltage during operation to implement a precise conversion between energy (unit: kWh) and capacity (unit: Ah). From the open-circuit voltage curve, it can be known that voltage is related to the magnitude of SOC. Therefore, in energy dispatch, SOC instead of battery voltage can be utilized as the argument of power function to approximate current, i.e. $I=\boldsymbol{g}(SOC,P)$. Considering the time resolution of a day-ahead dispatch, it is assumed that,
\begin{assumption}
	\label{asp:I_0}
	EVB current remains stable between two adjacent decision points, and the current is equal to the initial value $ I_0 $ at the starting point of the decision step.
\end{assumption}
	
Denote the sampling interval by $ \delta t $ and decision step by $ \Delta t $, where $ \Delta t = n\delta t $. 
From the assumption, SOC takes the initial value at $ i=1 $ in a single decision step and power takes the average value ($ \Delta E/\Delta t $) during the current approximation. Denote the change of energy in a single decision step by $ \Delta E $ (unit: Wh), and the change of battery charge (unit: Ah) is:
\begin{equation}
\begin{aligned}
\label{eqa:SOC-DeltaC}
\Delta C &= \sum_{i=1}^n I(i)\delta t \approx \boldsymbol{g}  \left( SOC_0,\frac{\Delta E}{\Delta t}  \right) \Delta t
\end{aligned}	
\end{equation}
where $SOC_0$ is defined by the normalized average lithium-ions concentration of solid phase in the negative electrode $\xi_{s}^-$ \cite{gu2021simplified}, i.e. $SOC_0 = SOC(1) = \frac{\frac{\sum_{j=1}^{\nu_s}\xi_{s}^-(d_j,i=1)}{\nu_s} - \xi_{\mathrm{min}}^-}{\xi_\mathrm{max}^- -  \xi_\mathrm{min}^-}$.

Denote the average power $\Delta E/\Delta t$ by $P_0$, which is obtained by simulating the $\EM(\cdot)$ with $ SOC_0 $ and $ I_0 $ from the feasible region given by SOPT as inputs. The uniform sampling inputs of $ SOC_0 $ and $ I_0 $ generate $P_0$ and their combination constitutes the power dynamics surface. Under fixed cell type (i.e. electrode materials) and ambient temperature, experimental evidence reveals that the power dynamics surface $ s_P(SOC_0,P_0,I_0)|_{\mathrm{Type},\mathrm{\Theta_{amb}}}  = 0 $ can be fitted as a plane, and current estimation $ \boldsymbol{g} $ equals the inverse function of $s_P$:
\begin{equation}
\label{eqa:SOC-I_0}
I_0 \approx \boldsymbol{g}(SOC_0,\frac{\Delta E}{\Delta t})\approx a_0 + a_1 SOC_0 + a_2 P_0
\end{equation}
where the first approximation stems from the definition of $ P(k) $ and the second from fitting the coefficients $ a_{0,1,2} $. Denote the battery cell capacity by $ C_0 $, thus SOC in dispatch can be updated by (\ref{eqa:SOC-DeltaC}),(\ref{eqa:SOC-I_0}):
\begin{equation}
\label{eqa:SOC-Update-cell}
\Delta SOC = \frac{-\Delta C}{C_0} \approx \frac{-I_0 \Delta t}{C_0}
\end{equation}

\subsubsection{Heat dynamics}
\label{sec:Heat dynamics}
Dynamic cell temperature of EVB affects the available power performance during dispatch. In a fast charging scenario, the internal heat generation is significant enough to change the cell temperature, while the external heat transfer is also notable. 

Therefore, it is necessary to regard the cell temperature as one of the battery states and estimate it during the optimization.
Considering the model complexity, a lumped heat model (\ref{eqa:Thermal_model}) from \cite{bizeray_lithium-ion_2015} is sufficient for the research requirement to approximate the temperature variation:
\begin{equation}
\label{eqa:Thermal_model}
mC_p\frac{\delta \Theta(t)}{dt} = H_i(t) + h_cA_{surf}(\Theta_{amb}(t) - \Theta(t)) + h_eH_e(t)
\end{equation}
where $m$ is the cell mass, $C_p$ is the heat capacity of battery cell, and the left side of (\ref{eqa:Thermal_model}) is the derivative of cell temperature. On the right side of (\ref{eqa:Thermal_model}), the first term $H_i$ represents the net heat generation in the cell due to electro-chemical energy conversion. $h_c$ and $h_e$ are lumped heat transfer coefficients, $H_e(t)$ denotes the net heat generation from other pack components. The second term and the third term on the right side are the heat exchange with the external environment and heat transfer of non-ideal pack components respectively.

In order to establish a connection of EVB temperature updates between the second-level EM model and the minute-level optimization, heat dynamics essentially depict the cross-sectional thermal characteristics of the EM simulation after one decision step. A lumped function $ \boldsymbol{\kappa} $ is proposed to model the summation of discrete temperature variation after one decision step:
\begin{equation}
\label{eqa:kappa}
\Delta \Theta = \int_{t}^{t + \Delta t}\delta \Theta(t) \approx \sum_{i=1}^n \delta \Theta(i) =\boldsymbol{\kappa} (\Theta_0, P_0)
\end{equation}
where $ \Delta \Theta $ is the temperature variation between decision steps and $ \Theta_0 $ is the initial temperature of this decision step $t$. $ \delta \Theta $ can be obtained by solving (\ref{eqa:Thermal_model}) and the average power $ P_0 $ can be derived from the inverse function of $ \boldsymbol{g} $.

To estimate heat dynamics function $ \boldsymbol{\kappa} $, EM is simulated with a uniform sampling of $ \Theta_0 $ and $ I_0 $ bounded by feasible region as inputs. The experiment demonstrates that $ \boldsymbol{\kappa} $ can be fitted as two separate planes on the positive and negative intervals of $ P_0 $, i.e. discharge and charge power, respectively.
\begin{equation}
\label{eqa:Temp-Update-cell}
\Delta \Theta \approx \boldsymbol{\kappa} \approx 
\begin{cases}
	e_0 + e_1\Theta_0 + e_{2,\dis}P_0, & \text{if } P_0 \geq 0 \\
	e_0 + e_1\Theta_0 + e_{2,\charge}P_0, & \text{if } P_0 < 0 \\
\end{cases}
\end{equation}
where the first approximation stems from the definition of $ \Delta \Theta $ and the second from fitting.

\subsubsection{State of Power-Thermal}
\label{sec:SOP}
SOPT is defined as the maximum available power to be dispatched at the present states, while satisfying the operating requirements in EM simulation as well. In terms of reaction mechanism, SOC is related to open-circuit voltage and the potential of power output, while cell temperature influences the intensity of reaction. Thus both of them are represented as arguments in SOPT.

Following Assumption.~\ref{asp:I_0}, under fixed ambient temperature and material type, the current $I_0$ is the only independent variable of EM if $SOC_0$ and $\Theta_0$ are taken as arguments, so its maximum determines the maximum available power. Therefore, the SOPT can be estimated by a small-scale nonlinear non-convex optimization of current $I_0$, which can be directly solved by heuristic methods:
\begin{equation}
\begin{aligned}
	\max_{I_0}:   &\quad \left| I_0 \right| \\
	\text{subject to:} &\quad \EM(SOC_0, \Theta_0, \boldsymbol{I}, \boldsymbol{\psi}) \in \boldsymbol{\Omega},\quad I_0 \in \boldsymbol{\Omega}_I
\end{aligned}
\end{equation}
where $ \boldsymbol{\Omega} $ represents the set of EVB operating requirements, and $ \boldsymbol{\Omega}_I $ distinguishes the discharging and charging processes.

In $ \boldsymbol{\Omega} $, to ensure the EVB scheduling plan is physically operating feasible, constraints of cell voltage ($ \boldsymbol{V} $) and constraints of remaining-capacity-related Li-ion concentration ($ \boldsymbol{\xi}_{s}^{\pm} $) are included. To implement an efficient and safe dispatch plan, constraints of cell energy conversion efficiency ($ \boldsymbol{\eta} $) are introduced in $ \boldsymbol{\Omega} $, which helps to curtail the heat generation during energy conversion in non-ideal cells. To prevent excess EVB aging, aging-related solution-solid interface potential ($ \boldsymbol{\phi}_{se}^{-} $), and active lithium loss ($ \boldsymbol{\tau}_{L} $) are set as limitations as well.

By taking material type and ambient temperature as hyper settings, SOPT can be mapped by uniform sampling arguments of initial SOC and cell temperature. Denote the optimal current by $\hat{I}_0$ and the corresponding voltage output from EM by $\hat{\boldsymbol{V}}$, and the total energy during a decision step by $\Delta \hat{E}$. Since there exists no controlling space for optimization between two decision steps, regarding the conservation of energy, the maximum available power defined under time scale $\Delta t $ equals the average power defined under $\delta t$:
\begin{equation}
\mathrm{SOPT} \equiv \frac{\Delta \hat{E}}{\Delta t} = \frac{\sum_{i=1}^n \hat{p}(i)\delta t}{\Delta t} =  \frac{\hat{I}_0\sum_{i=1}^n \hat{V}(i)}{n} 
\end{equation}

Experiments reveal that all $ \rm{SOPT} $\textasciitilde$SOC_0$\textasciitilde$\Theta_0$ values constitute a convex hull that can be piece-wised by multiple planes, where SOPT is expressed in parameterized functions:
\begin{equation}
\label{eqa:SOP-Est}
\begin{aligned}
	\boldsymbol{f}_d(SOC_0, \Theta_0)=& \min(b_{2,\dis}^m SOC_0+b_{1,\dis}^m \Theta_0 +b_{0,\dis}^m)\\
	\boldsymbol{f}_c(SOC_0, \Theta_0)=& \max(b_{2,\charge}^m SOC_0+b_{1,\charge}^m\Theta_0+b_{0,\charge}^m)\\
\end{aligned}
\end{equation}
where $m=1,\dots,M$ and $ M $ denotes the number of piece-wise segments. $ \boldsymbol{f}_d $ and $ \boldsymbol{f}_c $ denote the SOPT function during discharging and charging determined by initial SOC and cell temperature. $b_0^m, b_1^m, b_2^m$ are the fitted coefficients. It should be noted that the coefficients in power dynamics, heat dynamics, and SOPT are defined and fitted under a fixed material type denoted by $\mathrm{Type}$ and ambient temperature $\Theta_{amb}$, while different settings correspond to different sets of fitted parameters.

\subsection{Integration of LPC with Dispatch}
\subsubsection{Recursive Constraints}
\begin{figure}[htb]
\centering
\includegraphics[width=0.8\linewidth]{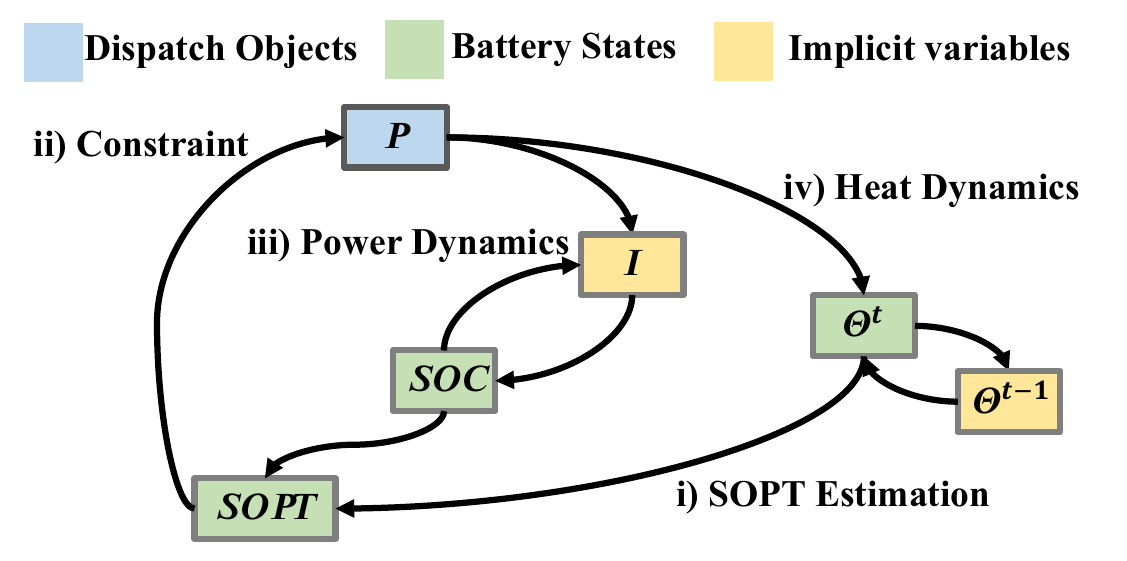}
\caption{Recursive EVB Constraints}
\label{fig:ConsRelation_Review}
\end{figure}

With the linearized LPC results, performance constraints of an EVB can be given recursively as Fig.~\ref{fig:ConsRelation_Review}. With these recursive constraints, the dispatch problem can be represented as an elementary MILP problem (M1) instead of the hierarchical nonlinear form M0.

As shown in Fig.~\ref{fig:ConsRelation_Review}, EVB power is the decision variable to be scheduled, while EVB states are updated with LPC results.
\begin{enumerate}[label=\itshape\roman*\upshape)]
\item \label{item:Cons_a} SOPT is estimated by (\ref{eqa:SOP-Est}), where $SOC_0$ equals to $SOC^t$ updated by last step and initial $\Theta_0$ equals to $\Theta^t$;
\item EVB power subjects to the convex feasible region formed by both discharging and charging SOPT;
\item According to power dynamics in section \ref{sec:Power dynamics}, both EVB power and SOC produces the magnitude of current $ I^t $. Current updates SOC iteratively, but in (\ref{eqa:SOC-Update}), vectorization enables non-iterative SOC update without the explicit expression of current.
\item \label{item:Cons_d} Cell temperature changes according to heat dynamics in section \ref{sec:Heat dynamics} during each step, thus the sequential cell temperature is updated by EVB power and itself, where the latter one is an implicit variable in (\ref{eqa:Temp-Update}) as well.
\end{enumerate}
For $ \forall t \in [1,T-1] $, \ref{item:Cons_a} - \ref{item:Cons_d} constitute recursive constraints.

\subsubsection{Matrix-based state update}
When applying power dynamics in the MILP optimization problem, iterative updates of sequentially coupled states of each EVB cause intricate variable correlations and consume a considerable amount of time during model formulation and regularization. 

Thus in (\ref{eqa:SOC-Update}) we propose a non-iterative matrix-based SOC update method with power dynamics. It is a collective update function for the entire problem, covering decision steps ranging from $ \tS $ to $ \tE $. The first term of (\ref{eqa:SOC-Update}) is the descending effect of initialization as time passes, while the second term is the summing effect of EVB power at present and prior to this moment. The third term is the fundamental effect such as the self-discharge, which is represented by $ a_0 $.

Moreover, the updating process avoids the need to explicitly compute the intermediate variable of current in (\ref{eqa:SOC-I_0}), so as to reduce the number of variables.
\begin{equation}
\label{eqa:SOC-Update}
\boldsymbol{SOC} = SOC_{\init} * \boldsymbol{A}  + \frac{1}{N_{\LC}} * \boldsymbol{B}_P(\boldsymbol{P}_{\LC,d} + \boldsymbol{P}_{\LC,c}) + \boldsymbol{B}_{\CS}
\end{equation}
\begin{equation}
\label{eqa:A_mat}
\chi = 1 - \frac{a_1}{C_{0,\LC}\Delta t},
~\boldsymbol{A} = \begin{bmatrix}
	\chi & \chi^2 & \cdots & \chi^T
\end{bmatrix}^{\mathrm{T}}
\end{equation}
\begin{equation}
\label{eqa:B_mat}
\begin{gathered}
	\boldsymbol{B} = \begin{bmatrix}
		1 & 0 & \cdots & 0 \\
		\chi & 1 & \cdots & 0 \\
		\vdots & \vdots & \ddots & \vdots \\
		\chi^{T-1} & \chi^{T-2} & \cdots & 1
	\end{bmatrix} \\
	\boldsymbol{B}_P = -\frac{a_2 \Delta t}{C_{0,\LC}} * \boldsymbol{B},
	~\boldsymbol{B}_{\CS} = - \frac{a_0 \Delta t}{C_{0,\LC}} * \boldsymbol{B} * \mathbf{1}_T
\end{gathered}
\end{equation}
where the LiB category is denoted by $\LC$ ($\LC=\ES/\EV/\BSS$), and $SOC_{\init}$ is the initialization of LiB before optimization. $\boldsymbol{SOC}$ is a vector, consisting of $SOC^t$ that is defined for each decision step. Similarly,  $\boldsymbol{P}_{\LC,d}$ and $\boldsymbol{P}_{\LC,c}$ represent discharge and charge power vector, respectively. $ N_{\LC} $ denotes the number of cells in a EVB, and $ \mathbf{1}_T $ is a vector of ones with a length of $T$.

Besides, with $\kappa$, the temperature in dispatch is updated by $ \Theta^{t+1} =  \Theta^t + \boldsymbol{\kappa} (\Theta^t, P^t_d + P^t_c)$. To apply heat dynamics without iterative updates, a similar matrix-based temperature update (\ref{eqa:Temp-Update})-(\ref{eqa:D_mat}) can be derived.
\begin{equation}
\label{eqa:Temp-Update}
\boldsymbol{\Theta} = \Theta_{\init} * \boldsymbol{C}  + \frac{1}{N_{\LC}} * (\boldsymbol{D}_{P,\dis}\boldsymbol{P}_{\LC,d} + \boldsymbol{D}_{P,\charge}\boldsymbol{P}_{\LC,c}) + \boldsymbol{D}_{\CS}
\end{equation}
\begin{equation}
\label{eqa:C_mat}
~\boldsymbol{C} = \begin{bmatrix}
	1 + e_1 & (1 + e_1)^2 & \cdots & (1 + e_1)^T
\end{bmatrix}^{\mathrm{T}}
\end{equation}
\begin{equation}
\label{eqa:D_mat}
\begin{gathered}
	\boldsymbol{D} = \begin{bmatrix}
		1 & 0 & \cdots & 0 \\
		1 + e_1 & 1 & \cdots & 0 \\
		\vdots & \vdots & \ddots & \vdots \\
		(1 + e_1)^{T-1} & (1 + e_1)^{T-2} & \cdots & 1
	\end{bmatrix} \\
	\boldsymbol{D}_{P,\dis} = e_{2,\dis} * \boldsymbol{D},
	~\boldsymbol{D}_{P,\charge} = e_{2,\charge} * \boldsymbol{D},
	~\boldsymbol{D}_{\CS} = e_0 * \boldsymbol{D} * \mathbf{1}_T
\end{gathered}
\end{equation}
where $\Theta_{\init}$ is the initial temperature of the entire optimization and other notations remain the same as those in the SOC update.

Equation (\ref{eqa:SOP-Est}) gives SOPT value. By incorporating the vectorization of $ \boldsymbol{SOC} $ in (\ref{eqa:SOC-Update}) and $ \boldsymbol{\Theta} $ in (\ref{eqa:Temp-Update}), SOPT can be expressed in matrix-based non-iterative form as well.

\section{Grid-connected EV Dispatch Considering EM}
\label{sec:DispatchModel}
\subsection{The Charging EV Dispatch}
\label{sec:operation-EV}
\subsubsection{Vehicle behavioral analysis}
The temporal characteristics of V2G participation are related to EVs' states of plug in. Considering a personally-owned EV operation in one day, it is divided into three different states: 
\begin{enumerate*}[label=\itshape\alph*\upshape)]
\item Plug in at a commercial building ($S_{\EV,C}^{j_g,g,t} = 1$), where an assorted bi-directional V2G-equipped CS is installed.
\item Plug in at residential building ($S_{\EV,R}^{j_g,g,t} = 1$), where EVs are connected to uni-directional V2G charging piles as a adjustable load.
\item Unplug when cruising ($S_{\EV,D}^{j_g,g,t} = 1$), where EVs only discharge.
\end{enumerate*}
 
EVs are divided into several groups $ g $ to demonstrate different living habits of vehicle owners, which lead to varied features of EV time profile. According to the definition, EVs are connected to the grid when owners are at work or home ($S_{\EV,C}^{j_g,g,t} = 1$, $S_{\EV,R}^{j_g,g,t} = 1$), complemented by unplug state between the above two grid-connected states. 

\subsubsection{Day-ahead dispatch of charging EVs}
Denote the group number of EV by $g$, the serial number of EV by $j_g$, and each group contains $n_g$ EVs with $j_g = 1,...,n_g$.

\paragraph{EV temporal characteristic}
Denote each EV's arrival and departure time of commercial building by $t_{\arrive,\alpha}^{j_g,g}, t_{\leave,\alpha}^{j_g,g}$, which are randomly sampled from the parametric intervals distinguished by group $[T_{\arrive,1}^{g}, T_{\arrive,2}^{g}], [T_{\leave,1}^{g}, T_{\leave,2}^{g}]$. EVs may arrive at or leave the commercial building multiple times, denoted by $\alpha$. Thus the state of plug at a commercial building, $S_{\EV,C}^{j_g,g,t} = 1$ when $t\in T_C=\{t\mid t_{\arrive,\alpha}^{j_g,g},...,t_{\leave,\alpha}^{j_g,g}\},\forall \alpha$. Denote the randomly parametric length of cruising time by $t_{L_D}^{j_g,g}$, and the state of unplug is intermediate between the state of plug in at the residential building or the commercial building, thus $S_{\EV,D}^{j_g,g,t} = 1$ when $t\in T_D=\{t\mid t_{\arrive,\alpha}^{j_g,g}-t_{L_D}^{j_g,g},...,t_{\arrive,\alpha}^{j_g,g}-1\}\cup \{t\mid t_{\leave,\alpha}^{j_g,g}+1,t_{\leave,\alpha}^{j_g,g}+t_{L_D}^{j_g,g}\}, \forall \alpha$. EVs are located at the residential building when owners are not working or driving, thus the state of plug in at the residential building, $S_{\EV,R}^{j_g,g,t} = 1$ when $t\notin (T_C \cup T_D)$.

\paragraph{Optimal EV charging with EM}
The initialization of EVB SOC is set by parameters and the initial EVB cell temperature equals ambient temperature. Incorporating grid-side decision variables, EVB SOC $SOC_{\EV}^{j_g,g,t}$ and cell temperature $\Theta_{\EV}^{j_g,g,t}$can be updated with (\ref{eqa:SOC-Update}),(\ref{eqa:Temp-Update}) by the sequential vector of discharging and charging power of each cell, where $ P_0 \equiv \Delta E/\Delta t $ is the power on each cell:
\begin{equation}
\label{eqa:EV-I}
\begin{gathered}
	P_{0,EV}^{j_g,g,t} = \frac{P_{\EV,d}^{j_g,g,t} + P_{\EV,c}^{j_g,g,t} + P_{\EV,\unplug}^{j_g,g,t}}{N_{\EV}}\\
	SOC_{\EV}^{j_g,g,t=1} = SOC_{\EV,\init}^{j_g,g},
	\Theta_{\EV}^{j_g,g,t=1} = \Theta_{amb}
\end{gathered}	
\end{equation}
where $ SOC_{\EV,\init}^{j_g,g} $ is acquired from dataset and $ \Theta_{amb} $ denotes the ambient temperature.

At the residential building where uni-directional V2G exists, EVBs are limited to the charging SOPT constraints to ensure practical power injection. In addition to this, for bi-directional V2G at the commercial building, EVBs are limited to both charging and discharging SOPT constraints:
\begin{equation}
\label{eqa:EV-Dis}
\boldsymbol{f}_d(SOC_{\EV}^{j_g,g,t}, \Theta_{\EV}^{j_g,g,t}) N_{\EV} S_{\EV,C}^{j_g,g,t} \geq P_{\EV,d}^{j_g,g,t} \geq 0
\end{equation}
\begin{equation}
\label{eqa:EV-Char}
\boldsymbol{f}_c(SOC_{\EV}^{j_g,g,t}, \Theta_{\EV}^{j_g,g,t}) N_{\EV} (S_{\EV,C}^{j_g,g,t} + S_{\EV,R}^{j_g,g,t}) \leq P_{\EV,c}^{j_g,g,t} \leq 0
\end{equation}
where $ P_{\EV,d}^{j_g,g,t}, P_{\EV,c}^{j_g,g,t} $ denote the discharging and charging power respectively and $ N_{\EV} $ denotes cell amount in an EVB.

In unplug state ($ S_{\EV,D}^{j_g,g,t} = 1 $), the rate of EVB discharging is related to irrelevant factors of scheduling, e.g., road quality and owners' driving practices. Thus, it can be assumed that the average cruising power is proportional to SOPT, where the ratio parameters denoted by $ r^{j_g,g}_{\EV,D} $ of each EV are randomly sampled within a certain range. Therefore, the cruising power of unplug state can be derived from SOPT:
\begin{equation}
\label{eqa:EV-Unplug}
P_{\EV,\unplug}^{j_g,g,t} = r^{j_g,g}_{\EV,D} S_{\EV,D}^{j_g,g,t} \boldsymbol{f}_d(SOC_{\EV}^{j_g,g,t}, \Theta_{\EV}^{j_g,g,t})
\end{equation}

Considering the benefits of owners and to ensure the energy feasibility of driving, EVs should be adequately charged to SOC requirements before the departure from commercial building $ SOC_{\EV,\req }^{j_g,g} \leq SOC_{\EV}^{j_g,g,t=t_{\leave,\beta}} $, which will transfer part of the EVB energy cost towards the commercial building. In order to promote EV owners to participate in V2G and submit to day-ahead dispatch, the allowable upper and lower boundaries of EVB SOC should be more conservative to extend battery life $ [SOC_{\EV,\min}, SOC_{\EV,\max}] $.

\subsection{The Battery-Swapping EV Dispatch}
\label{sec:BSS-Model}
\subsubsection{The operation analysis of BSS}
BSS provides EV swapping services with full-charged batteries for fast energy replenishment while participating in bi-directional V2G with its in-station EVB stockings. The major form of BSS dispatch is MILP where integers designate the swapping logic and the main goal of the BSS schedule includes: 
\begin{enumerate*}[label=\itshape\alph*\upshape)]
\item Energy Control: Precise and efficient charging and discharging strategy of the in-station EVBs.
\item EVB Allocation: Profitable dispatch of EVBs and swapping decisions.
\end{enumerate*}

\subsubsection{The design of multi-Layer optimization model (M2)}
\label{sec:BSS_M2}
Incorporating LPC results with typical BSS dispatch models can be inefficient or even impractical in the case of large-scale BSS. Because in NP-hard MILP, the numbers of binary decision variables and EVB recursive constraints are mixed with each other, which are both proportional to BSS storing size. Thus, there exists a necessity to propose an enhanced dispatch model which utilizes the EM effectively and efficiently.

To address this issue, a multi-layer optimization model of BSS is proposed in this paper, denoted by M2. In M2, EVBs are virtually divided into online and offline categories, which represent whether the battery is operating (either charge/discharge) or not. A virtual warehouse layer contains all offline EVBs and separates the grid-connected layer and business layer. Only online EVBs in grid-connected layer are dispatched for providing V2G service, while offline EVBs in the virtual warehouse are assigned to retain two discrete SOC states (Full-Charged/Depleted, denoted by subscript f and e respectively). The dispatcher takes the SOPT and SOC of each online EVB but only quantities of offline EVBs into account.

The virtual warehouse serves as a buffer between the inconsistency of aggregated binary signal and single external demand as well. Besides, instead of swapping one for one like the swapping logic in \cite{sarker_optimal_2015}, the allocation of online EVBs in M2 is unified dispatched, i.e. the exits of full-charged batteries from CDs are independent of the entrance of depleted batteries. In such cases, the associated CDs are allowed to be occasionally unoccupied by EVB, which provides transitory balancing flexibility of V2G and swapping business.

The object of dispatching in M2 is CDs, which serve as the reference of artificially defined states including SOPT and SOC. These definitions are essentially the sequential combination of real states of the EVBs that occupy the corresponding CD, thus they are non-zeros only if positive occupying state. 

Additionally, BSS is usually equipped with advanced cooling systems to keep the temperature constant, so EVBs in BSS can be assumed to be operating under thermostatic conditions.

\subsubsection{The Consideration of EM in BSS}
$ k $ denotes the index of CD on the grid-connected layer in M2. CDs' occupation states by EVB $ s^{k,t} $ can be represented as step signals. And the positive occupying state ($ s^{k,t}=1 $) is defined that $ k^{\mathrm{th}} $ CD retains a dispatchable EVB that is available to participate in grid-connected bi-directional V2G at $ t $.

Indicating signals of CDs appear as unit pulses, including online signals ($ x_{\ON,f}^{k,t}, x_{\ON,e}^{k,t} $) and offline signals ($ x_{\OFF,f}^{k,t}, x_{\OFF,e}^{k,t} $). They are only positive at the decision step of EVB retrieval.

SOC referring to CDs will be updated based on EVB current if the positive occupying state, while remains zero if unoccupied. Denote $ SOC_{\BSS,\upd} $ as the continuous SOC updated as power dynamics (\ref{eqa:SOC-Update-cell}) with the manifestation in (\ref{eqa:BSS-I}). At the moment of EVB going online ($ x_{\ON,\all}^{k,t} = 1 $),  SOC referring to CDs should be relaxed as (\ref{eqa:M2-SOC-Update}) since EVB contains discrete SOC of target warehouse as (\ref{eqa:M2-SOC_Req}). Let $ \mathcal{M}_b $ be a large constant, and (\ref{eqa:M2-SOC-Update}) is the convex form of SOC update.
\begin{equation}
\label{eqa:BSS-I}
P_{0,\BSS}^{k,t} = \frac{P_{\BSS,d}^{k,t}+P_{\BSS,c}^{k,t}}{N_{\BSS}},
SOC_{0,\BSS}^{k,t} = SOC_{\BSS}^{k,t}
\end{equation}
\begin{equation}
\begin{aligned}
\label{eqa:M2-SOC-Update}
&-\mathcal{M}_b(1-s^{k,t}) \leq SOC_{\BSS, \temp}^{k,t+1} - SOC_{\BSS,\upd}^{k,t+1} \leq \mathcal{M}_b(1-s^{k,t})\\
&-\mathcal{M}_b(s^{k,t}+x_{\ON,\all}^{k,t}) \leq SOC_{\BSS, \temp}^{k,t+1} \leq \mathcal{M}_b(s^{k,t}+x_{\ON,\all}^{k,t})
\end{aligned}
\end{equation}

Moreover, it is assumed that the energy loss $ SOC_{BSS,l} $ exists during EVB stocking and replacement, which is equivalently taken into account at the moment of going online. Then the sequential SOC update defined towards CDs is:
\begin{equation}
\label{eqa:M2-SOC-Update-seq}
\begin{aligned}
SOC_{\BSS}^{k,t+1} &= SOC_{\BSS, \temp}^{k,t+1} 
+ x_{\ON,f}^{k,t}(SOC_{\BSS,F} - SOC_{\BSS,l}) \\
& + x_{\ON,e}^{k,t}(SOC_{\BSS,E} - SOC_{\BSS,l})
\end{aligned}
\end{equation}
where the big-M method should be applied to each term of (\ref{eqa:M2-SOC-Update-seq}) independently to avoid non-convex optimization. The boundary limitation $ [SOC_{\BSS,\min},SOC_{\BSS,\max}] $ and SOC initialization $ SOC_{\BSS,\init} $ are given by parameters.

According to the SOPT estimation (\ref{eqa:SOP-Est}) and the thermostatic condition, the BSS power is limited as:
\begin{equation}
\label{eqa:M2-Dis-Char}
\begin{gathered}
\boldsymbol{f}_d(SOC_{\BSS}^{k,t}, \Theta_{\BSS}) N_{\BSS} s^{k,t} \geq P_{\BSS,d}^{k,t} \geq 0,\\
\boldsymbol{f}_c(SOC_{\BSS}^{k,t}, \Theta_{\BSS})N_{\BSS}s^{k,t} \leq P_{\BSS,c}^{k,t} \leq 0 
\end{gathered}
\end{equation}
where $ \Theta_{\BSS} $ takes optimal operating temperature and (\ref{eqa:M2-Dis-Char}) should be rewritten as convex constraints as big-M method.

\subsubsection{Control Logic of EVB}
As the step functions can be expressed from the unit impulse, offline signals determine the reverse of the positive occupying state ($ s^{k,t}=1\rightarrow s^{k,t+1}=0 $) and vice-versa. Let $ x_{\ON,\all}^{k,t}, x_{\OFF,\all}^{k,t}$ respectively denote the integrated online and offline signal without distinguishing destination warehouse: $ x_{\sig,\all}^{k,t} = x_{\sig,f}^{k,t} + x_{\sig,e}^{k,t},~\sig=\ON,\OFF $. Thus the CD's state can be independently updated according to corresponding indicating signals:
\begin{equation}
s^{k,t+1} = s^{k,t} + x_{\ON,\all}^{k,t} - x_{\OFF,\all}^{k,t}
\end{equation}
where the occupying states of CDs are initialized by parameters obtained from data: $ s^{k,\tS} = s^{k}_{\init} $.

Offline EVBs are concentrated in virtual warehouses and retain two discrete SOC levels. At the moment of EVB going offline ($ x_{\OFF,\all} = 1 $), SOC of the corresponding charging dock satisfies warehouse requirements of retrieval: Offline EVB should be sufficiently charged to retain SOC greater than $ SOC_{\BSS,F} $ before entering warehouse of full-charged EVB, otherwise it will enter warehouse of depleted EVB if $ SOC_{\BSS,E} $ is sufficiently charged.

\begin{equation}
\begin{aligned}
\label{eqa:M2-SOC_Req}
SOC_{\BSS,F} - SOC_{\BSS}^{k,t+1} \leq \mathcal{M}_b(1-x_{\OFF,f})\\
SOC_{\BSS}^{k,t+1} - SOC_{\BSS,F} \leq \mathcal{M}_b(1-x_{\OFF,e})\\
SOC_{\BSS,E} - SOC_{\BSS}^{k,t+1} \leq \mathcal{M}_b(1-x_{\OFF,e})
\end{aligned}
\end{equation}

Denote the total number of required EVB on the business side to be swapped at $ t $ by $ Q_{\req}^t $. On the business side, full-charged EVBs substitute depleted EVBs at $ t $ while the amounts of stocking EVBs ($ Q_f^t,Q_e^t $) in the warehouse change accordingly. To the grid-connected side, online and offline signals direct the EVB switching process between the virtual warehouse and CDs accordingly. Thus the amounts of stockings are updated according to both the indicating signals of CDs and the swapped quantity:
\begin{equation}
Q_\wh^{t+1} = Q_\wh^t+\sum_{k}x_{\OFF,\wh}^{k,t} - \sum_{k}x_{\ON,\wh}^{k,t} \mp Q_{\req}^t,~\wh = f,e
\end{equation}

Besides, the amounts of stockings in the virtual warehouse are initialized by parameters $ Q_\wh^{\tS} = Q_{\wh,\init} $ and should be kept positive at all times $ Q_\wh^t \geq 0 $. 

To ensure the asset equilibrium and continuous operation of BSS over multi-days, the total stocking EVB amount is equal at the start and end points of the day as (\ref{eqa:M2-equilibrium}) while the amount is postulated to have coherence $ Q_\wh^{\tE} \in [Q_\wh^{\tS}-\epsilon, Q_\wh^{\tS}+\epsilon] $:
\begin{equation}
\label{eqa:M2-equilibrium}
Q_e^{\tS} + Q_{f}^{\tS} =  Q_e^{\tE} + Q_{f}^{\tE}
\end{equation}
where $ \epsilon $ denotes the minor allowed inconsistency of warehouse stock to coordinate with the aggregated method.

\subsubsection{Aggregation of Charging Dock}
To further enhance the solving efficiency, integer variables related to EVB control logic can be assembled chronologically and individually, which essentially suggests the aggregation of EVBs. This is reasonable because: 
\begin{enumerate*}
\item chronologically a lower update frequency for indicating signal is sufficient since the charging or discharging duration it takes to meet the warehouse layer's returning requirements is longer than normal decision step.
\item individually EVBs in each warehouse are assigned to identical SOC so they possess similar power characteristics when interacting with CDs.
\end{enumerate*}

Chronologically, let the integer decision variables annotated by subscript $ \ds $ have a longer update step $ \Delta h $ instead of $ \Delta t $. To coordinate the assembled integers with other normal continuous states like SOC and SOPT, and the high update frequency with the low ones, step signals have to retain the state while impulse signals only impact at the moment of state switch, i.e. the edge of step signals. As a result, a deterministic conversion between different resolutions is given (assuming $ \Delta h=\theta \Delta t $). Taking $x_{\ON,f,\ds}$ as the example:
\begin{equation}
\label{eqa:conv_s}
s_{\ds}^{k,h} = s^{k,\mathrm{t|t=\theta(h-1)+1:\theta h}}
\end{equation}
\begin{equation}
\label{eqa:conv_x}
x_{\ON,f,\ds}^{k,h}=1 \rightarrow 
\begin{cases}
x_{\ON,f}^{k,\mathrm{t|t=\theta (h-1)+1:\theta h-1}}=0\\
x_{\ON,f}^{k,\mathrm{t|t=\theta h}}=1
\end{cases}
\end{equation}
and $ x_{\ON,e,\ds}^{k,h}, x_{\OFF,f,\ds}^{k,h}, x_{\OFF,e,\ds}^{k,h} $ are similarly converted as (\ref{eqa:conv_x}).

Individually, multiple CDs can be controlled collectively with shared indicating signals under the consistency assumption due to the same SOC in each warehouse. SOC updates (\ref{eqa:M2-SOC-Update}) and SOPT constraints (\ref{eqa:M2-Dis-Char}) are still applicable, representing the power characteristics of a single EVB. But grid-side power injecting should be modified according to the assembled amount, which is considered in (\ref{eqa:Grid_C}).

Meanwhile, the amount update of warehouse stocking $Q_\wh$ is modified. Denote the amount of assembled CDs by $ \beta $:
\begin{equation}
\label{eqa:M2-Q-Update_h}
\begin{aligned}
Q_\wh^{h+1} &= Q_\wh^h + \beta\sum_{k}(x_{\OFF,\wh,\ds}^{k,h} - x_{\ON,\wh,\ds}^{k,h}) \mp Q_{\req}^h
\end{aligned}
\end{equation}

\section{Dispatch Model of Local Energy System}
\label{sec:LES}
\subsection{The Local Energy System (LES) Design}
In this article, an LES integrated with both CS and BSS is studied as the scenario of grid-connected EVs. High peak-shaving potential and renewables assimilation can be exploited from LES due to the energy storing capability provided by grid-connected EVs as the mobile storage as well as the stationary ES. 

In this LES, PV is installed on each residential building corresponding to each EV owner while the PV-ES is only located at the commercial building. Charging EVs travel between the residential and commercial buildings and park in the CS when they arrive at the commercial building as Section \ref{sec:operation-EV} stated. BSS modeled in Section~\ref{sec:BSS-Model} provides services for the public. Both CS and BSS are capable of participating in bi-directional V2G, whereas only uni-directional V2G is available at the residential building.

\subsection{The Implementation of Optimal Dispatch}
The stability of LES is assured by the external grid, which only injects power into LES through the real-time energy market.

For grid-connected EVBs, in addition to energy conversion efficiency $ \eta $, lumping efficiency of grid-connected V2G interface denoted by $ \gamma $ should be considered due to non-ideal charging dock facilities, e.g. converters, distribution transformers, cables, etc. While other power electronic characteristics of the V2G interface such as harmonics are neglected since they have little impact on the day-ahead stage.

\begin{assumption}
	The lumping efficiency of V2G interface is consistent and stable in day-ahead energy dispatch.
\end{assumption}

Besides, non-ideal interface efficiency spontaneously relaxes the mutually exclusive constraint of EVB charge and discharge power($ P_cP_d=0 $), because cost increment will be induced in the economic optimization objective if an exclusive power is applied instead of an equivalently unitary power, as proved in~\cite{li_sufficient_2016}. Since the power equilibrium, the purchasing power is:
\begin{equation}
	\label{eqa:Grid_C}
	\begin{aligned}
		P_{G,C}^t =-P_{L,C}^t+ P_{\PV,\real,C}^t+(P_{\ES,d}^t\gamma+\frac{P_{\ES,c}^t}{\gamma})+\sum_{ix}P_{\SL}^{ix,t}\\
		+\sum_{j_g,g}(P_{\EV,d}^{j_g,g,t}\gamma+\frac{P_{\EV,c}^{j_g,g,t}}{\gamma}S_{\EV,C}^{j_g,g,t})+\beta\sum_k(P_{\BSS,d}^{k,t}\gamma+\frac{P_{\BSS,c}^{k,t}}{\gamma})
	\end{aligned}
\end{equation}
\begin{equation}
	\begin{aligned}
		P_{G,R}^{j_g,g,t} =-P_{L,R}^{j_g,g,t}+ P_{\PV,\real,R}^{j_g,g,t}
		+\frac{P_{\EV,c}^{j_g,g,t}}{\gamma}S_{\EV,R}^{j_g,g,t}
	\end{aligned}
\end{equation}
where $ P_{G}^t $ denotes the real-time purchased power from the grid at either commercial (labeled with subscript $ C $) or residential building (labeled with subscript $ R $). $ P_{\PV,\real}^t $ denotes the substantially utilized photovoltaic power, and $ P_{L}^t $ denotes fixed local loads whereas $ P_{\SL}^{ix,t} $ denotes $ \mathrm{ix}^{th} $ adjustable local loads. 

Additionally, LES only purchases electricity from external grid $ P_{G,C}^t \leq 0, P_{G,R}^{j_g,g,t} \leq 0 $, whereas without the capability of reversely selling considering the practical connections of the high-voltage transmission line. Photovoltaic power $ P_{\PV,\real,C}^t, P_{\PV,\real,R}^{j_g,g,t} $ is limited to the physical irradiation amount $ P_{\PV,\ir,C}^t, P_{\PV,\ir,R}^{j_g,g,t} $ respectively. And the adjustable loads satisfy the total requiring energy $ E_{\SL}^{ix} = \sum_tP_{\SL}^{ix,t}\Delta t $ in a whole day but real-time load power is scheduled within the limited range $ [P_{\SL,\min}^{ix}, P_{\SL,\max}^{ix}] $.

The design of ES is a direct amplification of the single EVB cell, whose available power can be directly derived from (\ref{eqa:SOP-Est}) as (\ref{eqa:ES-Dis}) and SOC update from (\ref{eqa:SOC-Update}) with the manifestation in (\ref{eqa:ES-I}). The SOC boundary $ [SOC_{\ES,\min},SOC_{\ES,\max}] $ and initialization $ SOC_{\ES}^{t=1} $ is set by parameters. ES is usually operating under thermostatic conditions since the independent thermal managing equipment in the station. Denote the operating temperature of ES is $\Theta_{\ES}$, $ P_{\ES,d} $ and $ P_{\ES,c} $ are respectively the discharging and charging power which subject to the SOPT: 
\begin{equation}
	\label{eqa:ES-Dis}
	\boldsymbol{f}_d(SOC_{\ES}^{t}, \Theta_{\ES})\geq \frac{P_{\ES,d}^{t}}{N_{\ES}}\geq 0 \geq \frac{P_{\ES,c}^{t}}{N_{\ES}} \geq \boldsymbol{f}_c(SOC_{\ES}^{t}, \Theta_{\ES})
\end{equation}
\begin{equation}
	\label{eqa:ES-I}
	P_{0,\ES}^t = \frac{P_{\ES,d}^{t}+P_{\ES,c}^{t}}{N_{\ES}}, SOC_{0,\ES}^t = SOC_{\ES}^{t}
\end{equation}

The goal of LES scheduling is to minimize the overall energy expenses $ Z $, which can be divided into electricity cost $ Z_{\CST} $ and swapping revenue $ Z_{\REV} $, in addition to asset appreciation $ Z_{\APR} $ of the storing energy as penalty. 

Denote the commercial and residential real-time electricity price by $ V_C^t, V_R^t $, respectively. The electricity cost is:
\begin{equation}
	Z_{\CST,C} = \sum_t -P_{G,C}^t V_C^t \Delta t, \ Z_{\CST,R} = \sum_{j_g,g}\sum_t -P_{G,R}^{j_g,g,t} V_R^t \Delta t
\end{equation}

Denote the swapping service fee by $ V_s $, and the business revenue of BSS $ Z_{\REV} $ is related to the earnings per swapping $ V_{\BSS} $, which consist of electricity fee and service fee:
\begin{equation}
	\begin{aligned}
		V_{\BSS} = (SOC_{\BSS,F}-&SOC_{\BSS,E})C_{0,\BSS}N_{\BSS}U_{a,\BSS}V_a + V_s\\
		Z_{\REV} &= V_{\BSS}\sum_h Q_{\req}^h
	\end{aligned}
\end{equation}

The energy stored in a grid-connected EVB is regarded as an LES asset, which will induce appreciation if higher SOC at the end ($ \tE $) than the start ($ \tS $). Assume the LiB of ES and all charging EVs are in the LES scope at ($ \tE $) and ($ \tS $). Whereas in BSS, the SOC variation of EVBs on CDs and stocking change in the warehouse should be taken into account. Denote the average operating LiB voltage by $ \mathrm{U_a} $, and average electricity price in one day by $ \mathrm{V_a} $. Therefore, the asset appreciation $ Z_{\APR} $ can be approximated:
\begin{equation}
	\begin{aligned}
		&Z_{\APR,\LC} = \Delta SOC_{\LC}C_{0,\LC}N_{\LC}\mathrm{U_{a,\LC}}\mathrm{V_a},~\LC = \ES,\EV,\BSS\\
		&\Delta SOC_{ES} = (SOC_{\ES}^{\tE}-SOC_{\ES}^{\tS})\\
		&\Delta SOC_{EV}\sum_{j_g,g} (SOC_{\EV}^{j_g,g,\tE}-SOC_{\EV}^{j_g,g,\tS}), \\
		& \Delta SOC_{BSS} = 
		\begin{cases}
			M1:&\sum_k (SOC_{\BSS}^{k,\tE}-SOC_{\BSS}^{k,\tS}) \\
			M2:&\beta \sum_k (SOC_{\BSS}^{k,\tE}-SOC_{\BSS}^{k,\tS})\\ 
			& +((Q_f^{\tE}-Q_f^{\tS})SOC_{\BSS,F}\\
			& +(Q_e^{\tE}-Q_e^{\tS})SOC_{\BSS,L}
		\end{cases}
	\end{aligned}
\end{equation}

The overall optimization problem of LES dispatch is to acquire minimal energy expenses, while the charging EV model and BSS model are incorporated. M1 denotes the ordinary MILP in Appendix.~\ref{apx:BSS-M1}, while M2 is the proposed MILP dispatch considering EM efficiently with model enhancement as analyzed in the following remark.
\begin{equation}
	\begin{aligned}
		\label{eqa:Objective}
		&\begin{aligned}
			\text{(M2)~minimize:}   \quad Z &= Z_{\CST,C}+Z_{\CST,R}-Z_{\APR,\ES} \\&-Z_{\APR,\EV}-Z_{\APR,\BSS}-Z_{\REV} \\
		\end{aligned} \\
		&\text{subject to:}~(\ref{eqa:SOC-Update})-(\ref{eqa:EV-Unplug}),~(\ref{eqa:BSS-I})-(\ref{eqa:ES-I})
	\end{aligned}
\end{equation}

\begin{remark}
\label{sec:Analysis}
To utilize EM in optimal EVB energy dispatch, LPC is applied to extract generic constraints for EVB cells. Although such a characterization-based approach is applicable for linear programming problems, it is not sufficient in the grid-connected EV scenario, which is a large-scale MILP. The model becomes intricate for optimization, thus in the proposed dispatch model (M2), multiple techniques are applied to accelerate the optimizing process:
\begin{itemize}
	\item The iterative state update in sequentially coupled constraints is assembled in the matrix-form. LPC results are incorporated with the dispatch model without explicitly calculating redundant variables like current.
	\item The EVs' temporal distribution is grouped and profiled to demonstrate uncertain behaviors without extra integers. The available V2G potential of CS is determined by the EVs' state and their recursive constraints accordingly.
	\item A virtual warehouse layer in BSS is proposed to aggregate the EVBs that do not have energy transfer with the grid. In the warehouse layer, EVBs retain two discrete SOC, and other state variables are eliminated. 
	\item The dispatch objects in BSS are CDs instead of each EVB to control the number of decision variables. The states in optimization are artificially defined towards CDs, which are essentially sequential combinations of real states.
	\item The operating logic of BSS is assembled according to the consistency of EVBs' characteristics in the virtual warehouse layer and the lower update frequency of indicating signals than other normal continuous variables.
\end{itemize}
Compared to ordinary dispatch, the proposed model is effectively condensed and simplified for day-ahead application to enhance solution efficiency.
\end{remark}

\section{Case Study}
\label{sec:case-study}
\subsection{Parameters}
\subsubsection{Configuration of LiBs and LPC results}
As for the configuration of studied battery, the LiB settings take the value from \cite{gu2022sop}. The internal state restrictions $ \boldsymbol{\Omega} $ control the range of SOPT, i.e. the utilizing intensity of EVB. For instance, a higher energy conversion efficiency limitation leads to tighter power constraints and a conservative dispatch plan, but EVBs' heat generation and degradation are curtailed at the same time.

\begin{figure}[htbp]
	\centering
	\begin{minipage}{0.24\textwidth}
		\centering
		\includegraphics[width=\linewidth]{Figure//PowerDynamics.pdf}
		\subcaption{{Power Dynamics}}
		\label{fig:PowerDynamics}
	\end{minipage}
	\hfill
	\begin{minipage}{0.24\textwidth}
		\centering
		\includegraphics[width=\linewidth]{Figure//HeatDynamics.pdf}
		\subcaption{Heat Dynamics}
		\label{fig:HeatDynamics}
	\end{minipage}
	\begin{minipage}{0.24\textwidth}
		\centering
		\includegraphics[width=\linewidth]{Figure//SOPT.pdf}
		\subcaption{SOPT (Discharge)}
		\label{fig:SOPT}
	\end{minipage}
	\hfill
	\begin{minipage}{0.24\textwidth}
		\centering
		\includegraphics[width=\linewidth]{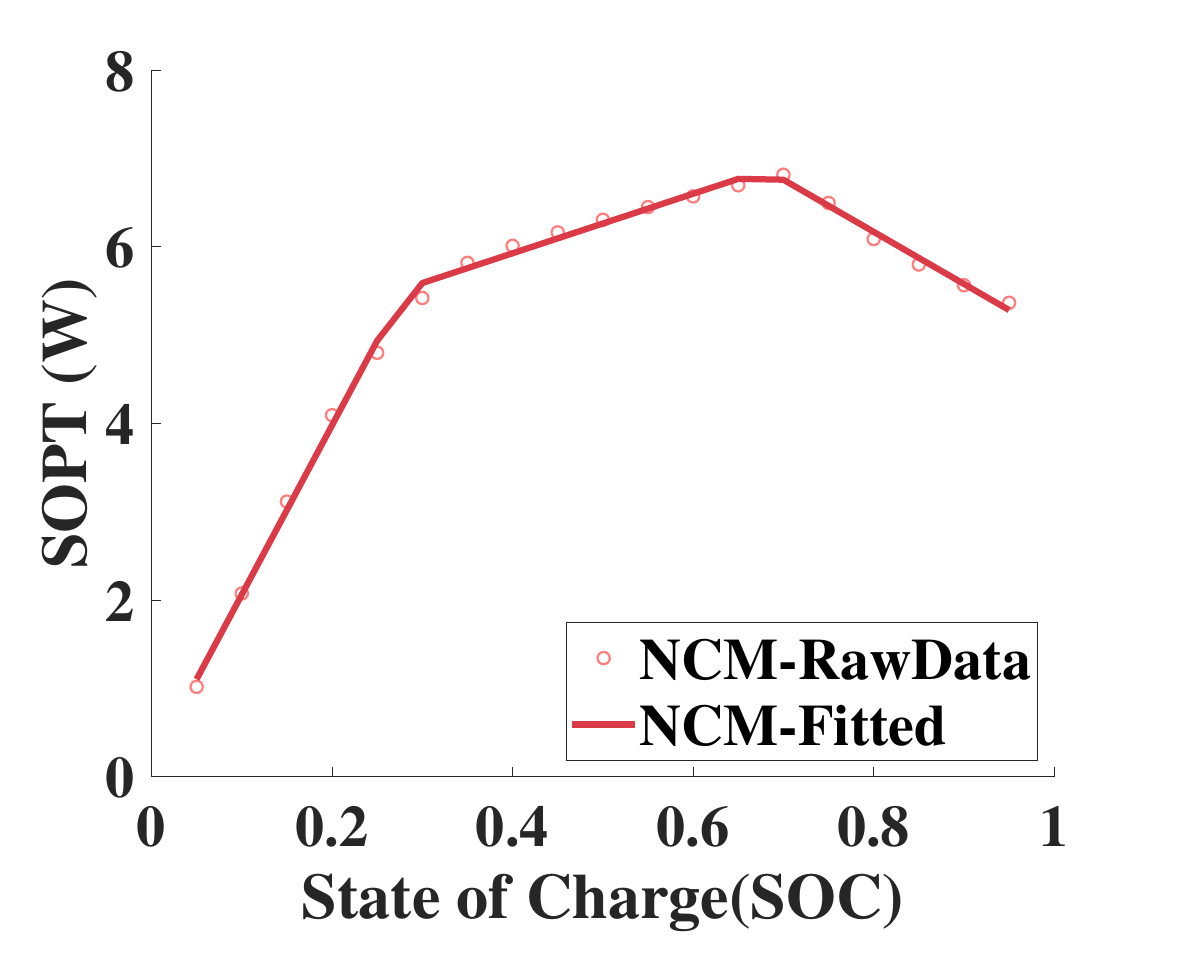}
		\subcaption{SOPT (Discharge, Thermostatic)}
		\label{fig:SOPT_line}
	\end{minipage}
	\caption{The LPC Results (NCM,25$^{\circ}$C)}
	\label{fig:LPC}
\end{figure}

Fig.~\ref{fig:LPC} reports the LPC results under 25$^{\circ}$C ambient temperature on NCM cells. Linearization results of LPC are summarized as fitting parameters ($ a_{0,1,2}(M=3),~b_{0,1},~e_{0,1,2}$). The LPC results are exactly determined by cell types, ambient temperature of LiBs, and other experimental settings including expected operating restrictions $\boldsymbol{\Omega}$. LPC results of LiBs that are composed of other cell types and LiBs under other ambient temperatures can be similarly given.

It should be noted that in Fig.~\ref{fig:HeatDynamics} the heat dynamics are fitted symmetrically regarding EVB power by two crossover planes. But in dispatch, due to the exclusive constraint $ P_c P_d = 0 $, the ``V"-style linearization is convex as the sum of Z-axis value mapped by $P_c$ and $P_d$ separately on the two planes.

In Fig.~\ref{fig:LPC}, it can be seen that the linearized results fit the raw data well. Moreover, goodness of fit ($ R^2 $) is used to validate the efficacy of applying linearization on LPC result. Table.~\ref{tab:L-Validation} reports the $ R^2 $ values which are all above 0.96, representing a generally accurate fit to be utilized in energy dispatch.

Additionally, to balance the computational complexity of EM simulation in LPC and dispatch optimization as stated in \ref{sec:PCTheory}, $ n = 900 $ samplings with $ \delta t = 1 ~\mathrm{s} $ interval are selected for the EM simulation, providing power characteristics applicable for the $ \Delta t = 15 ~\mathrm{min} $ day-ahead optimization. 

\begin{table}[htb]
	\centering
	\caption{Linearization Validation: Goodness of Fit}
	\begin{tabular}{ccc}
		\hline
		\textbf{}  & \textbf{Power Dynamics} & \textbf{Heat Dynamics}  \\  \hline
		\textbf{$ R^2 $}   & 0.99         & 0.99                \\ \hline
		\textbf{}  & \textbf{SOPT Estimation} & \textbf{SOPT Estimation (Thermostatic)} \\ \hline
		\textbf{$ R^2 $}   & 0.96         & 0.98                \\ \hline
	\end{tabular}
	\label{tab:L-Validation}
\end{table}

\subsubsection{LES setting}
The system-level settings are exhibited in this section. Considering the mainstream of manufacturing status nowadays, the LiBs for different applications are assigned with two electrode materials respectively as given in Table.~\ref{tab:LiB-Setting}. Besides, in the automotive sale sector, the capacity of LiB is usually identified in energy units, and the number of cells in a LiB module can be determined according to average voltage: $ \mathrm{N_{ST} = \frac{E_{ST}} {C_{0,ST}U_{a,ST}}}$. Regarding the charge/discharge depth and the consideration of private and non-private property, the SOC range is limited separately.

Table.~\ref{tab:LES-General} exhibits the setting of LES. The lumping efficiency of grid-connected V2G interface $ \gamma $ is assumed to be 0.85 considering practical facilities. EV discharging power ratio during unplug ($ r_{EV,D} $) is randomly sampled within the assumed interval. The total charging EV amount in LES is set to 100 and the number of grid-connected CDs supporting V2G in BSS is set to 25, which is aggregated on $ \beta $. Thus in addition to the stock in the warehouse, the total dispatched EVB located in BSS reaches 750. 
\begin{table}[htb]
	\centering
	\caption{Configuration of LiBs and Macro Settings}
	\label{tab:LiB-Setting}
	\begin{threeparttable}
		\resizebox{\columnwidth}{!}{%
			\begin{tabular}{cccccc}
				\hline
				& EV                    & BSS        & ES 		   & $C_0$ (Ah)      & $U_a$ (V) \\ \hline
				NCM  & \begin{tabular}[c]{@{}c@{}}[0.2,0.8]\tnote{*} \\ 35 kWh\end{tabular}  &  \begin{tabular}[c]{@{}c@{}}[0.1,0.9] \\ 50 kWh\end{tabular}     &    	/	   & 1.7572 			 & 3.43V 	 \\
				LFP  &/&/& \begin{tabular}[c]{@{}c@{}}[0.01,0.99]  \\ 1000 kWh\end{tabular} & 1.2184 			 & 2.86V 	 \\ \hline
			\end{tabular}%
		}
	\end{threeparttable}
	\begin{tablenotes}
		\footnotesize
		\item{*}: Specify the electrode material of LiB and the SOC range limitation
	\end{tablenotes}
\end{table}

\begin{table}[htb]
	\centering
	\caption{General Setting of LES Scheduling}
	\label{tab:LES-General}
	\resizebox{\columnwidth}{!}{%
		\begin{tabular}{ccccccc}
			\toprule
			\multicolumn{2}{c}{\begin{tabular}[c]{@{}c@{}}Decision   \\ Step (min)\end{tabular}} &
			\multirow{2}{*}{\begin{tabular}[c]{@{}c@{}}Ambient \\ Temperature\end{tabular}} &
			\multirow{2}{*}{T} &
			\multirow{2}{*}{$\gamma$} &
			\multirow{2}{*}{$r_{\EV,D}$} &
			\multirow{2}{*}{$SOC_{\EV,\req}$} \\ \cline{1-2}
			$\Delta t$      & $\Delta h$      &               &               &         &              &              \\ \hline
			15              & 60              & 25            & 96            & 0.85    & 0.15-0.30    & 0.7          \\ \midrule
			\multicolumn{2}{c}{$SOC_{\BSS,E}$} & $SOC_{\BSS,F}$ & $SOC_{\BSS,l}$ & $\beta$ & $Q_{e,\init}$ & $Q_{f,\init}$ \\ \hline
			\multicolumn{2}{c}{0.1}           & 0.9           & 0.02          & 6       & 300          & 300          \\ \bottomrule
		\end{tabular}%
	}
\end{table}

\begin{figure}[htb] 
	\centering
	\begin{subfigure}[t]{0.48\linewidth}{
			\begin{minipage}[t]{1\linewidth}
				\centering
				\includegraphics[width = 1\linewidth]{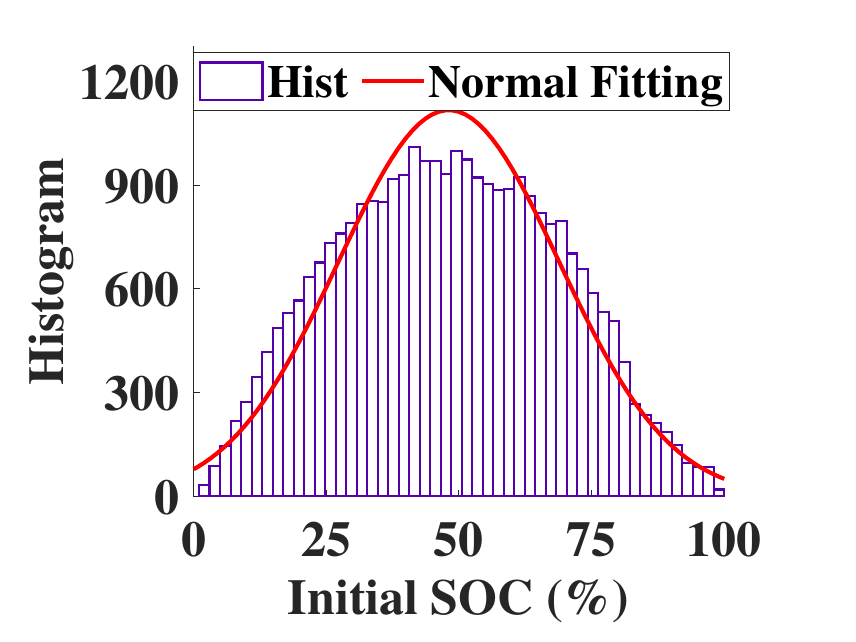}
				\subcaption{EV Initial SOC}\vspace{6pt}
				\label{fig:Data-Init-a}
		\end{minipage}}
	\end{subfigure}
	\begin{subfigure}[t]{0.48\linewidth}{
			\begin{minipage}[t]{1\linewidth}
				\centering
				\includegraphics[width = 1\linewidth]{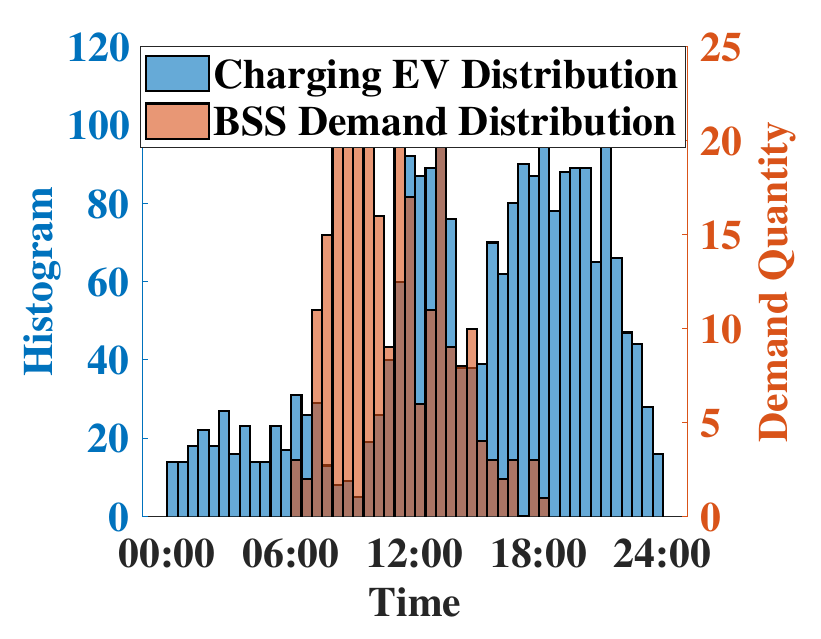}
				\subcaption{EV Distribution and BSS Demand}\vspace{6pt}
				\label{fig:Data-Init-b}
		\end{minipage}}
	\end{subfigure}
	\centering
	\caption{Input Data of EV and BSS}
	\label{fig:Data-Init}
\end{figure}

We acquire data of fundamental load and PV output on a typical day from the CoSSMic dataset\cite{data_2020}. While the real-time electricity price is from a random day in PJM Market.

The initialization of EV refers to the real operating data from several commercial CS in Shenzhen China. We investigate the initial SOC and the temporal distribution when EVs begin to recharge, as demonstrated in Fig.~\ref{fig:Data-Init-a},\ref{fig:Data-Init-b}. It can be seen that the initial SOC of charging EV may be approximated by the normal distribution centered at $ SOC=0.5 $. And the arrival time of charging EVs may mainly attribute to the owners' working routine. Thus, 4 groups of EVs are decomposed considering the mainstream of working routines, including day shift, night shift, driving home at midday, and frequent random charging (e.g., taxi). The number of day-shift EVs is twice as many as others, while the rest are evenly divided among other groups. Since there exists few available operating data of BSS, we apply the number of EVs entering a CS located at an office building provided in \cite{acndata_2019} to approximate BSS swapping demand, where the total demand quantity is scaled to be compatible with LES load data, as shown in Fig.~\ref{fig:Data-Init-b}.

\subsection{Day-Ahead Energy Dispatch Result}
\subsubsection{Case I: Thermostatic scenario}
In order to study the patterns of dispatch plans considering EM and the feature of grid-connected EVs when participating in V2G, we first conduct the case study under the moderate charging scheme, in which the EVBs are limited to operating with a conservative dispatch plan. The cell energy conversion efficiency constraint is set to 99.5\%, corresponding to a relatively moderate power constraint (\textasciitilde0.4C) and a curtailed heat generation due to less energy dissipated during the energy conversion. As a result, it is validated that under this scenario, the cell temperature variation is not significant ($ \Delta \Theta < 3 ^{\circ}C $) to affect SOPT when implementing the dispatch plan, so it is reasonable to assume that Case I is thermostatic to apply a better fit of LPC results.

\begin{figure*}[htb] 
	\centering
	\begin{subfigure}[t]{0.32\linewidth}
		\centering
		\includegraphics[width = 1\textwidth]{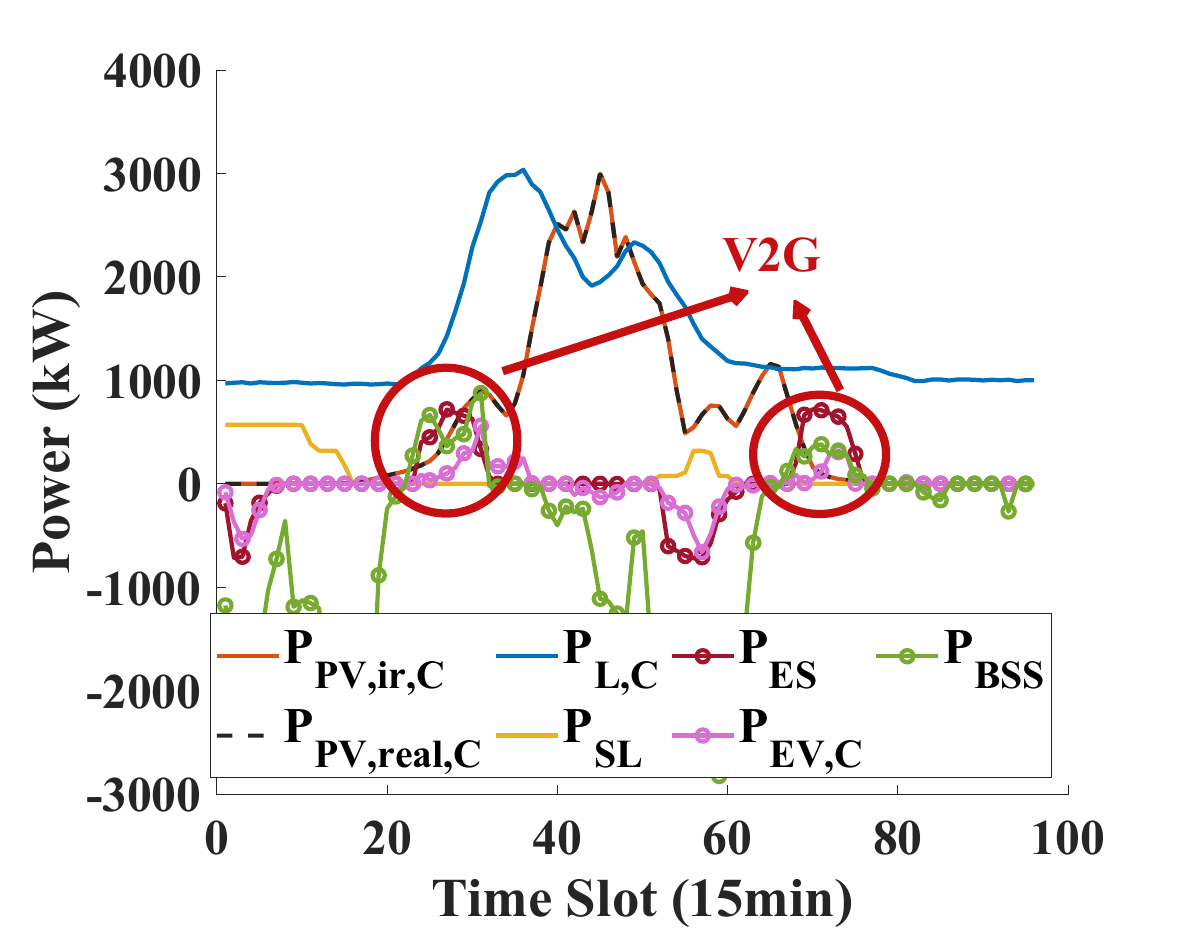}
		\subcaption{Power Dispatch Scheme}
		\label{fig:PowerScheme-a}
	\end{subfigure}
	\begin{subfigure}[t]{0.32\linewidth}
		\centering
		\includegraphics[width = 1\textwidth]{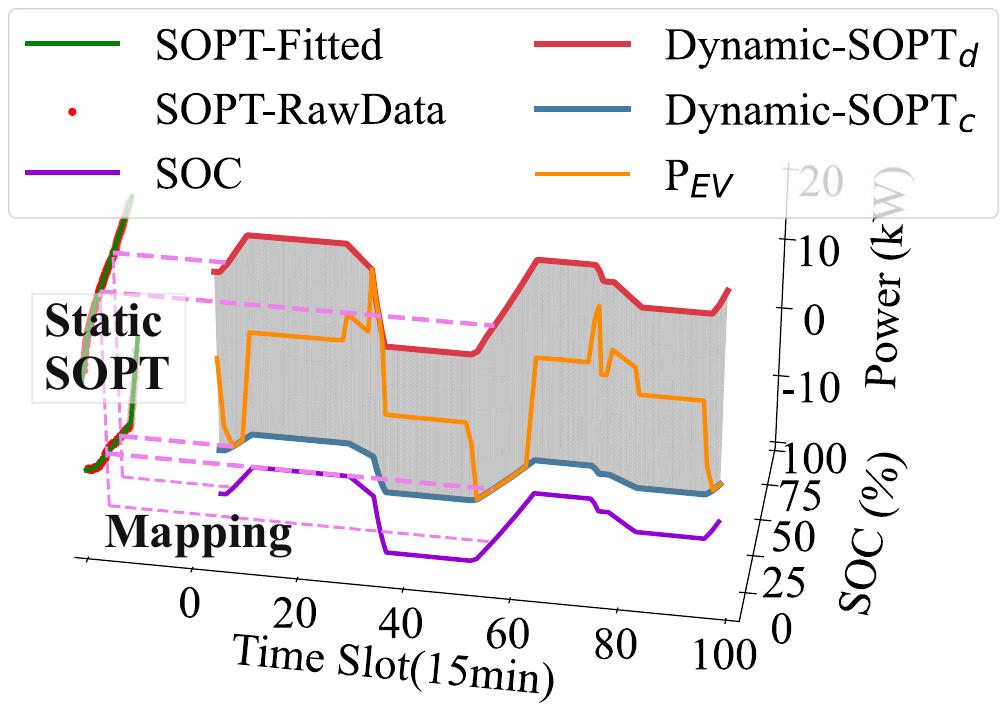}
		\subcaption{The Dynamic Available Power of EVB}
		\label{fig:EVResult-a}
	\end{subfigure}
	\begin{subfigure}[t]{0.32\linewidth}{
			\centering
			\includegraphics[width = 1\linewidth]{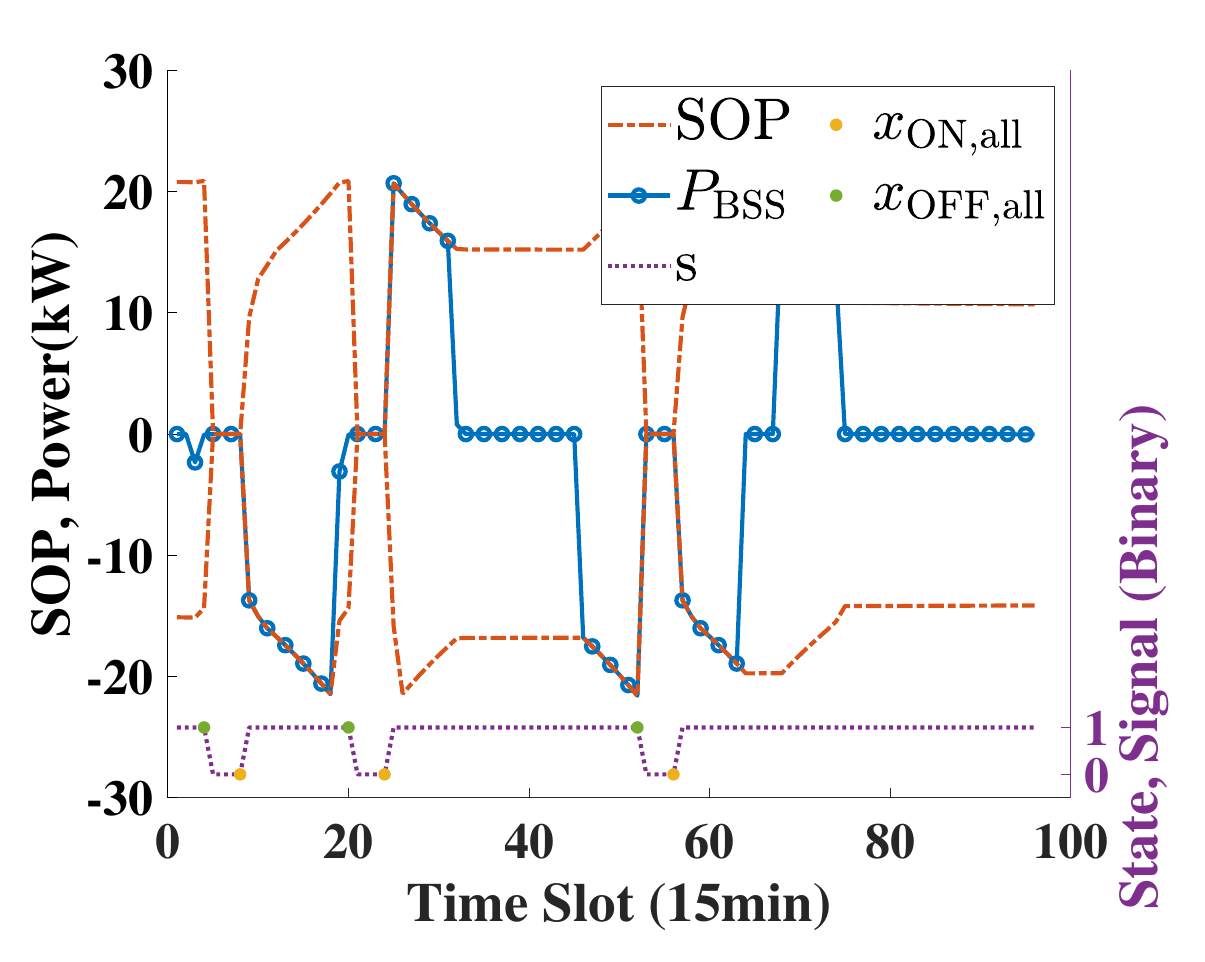}
			\subcaption{A Single EVB Power in BSS}\vspace{6pt}
			\label{fig:BSSResult-a}}
	\end{subfigure}
	\centering
	\caption{The Dispatch Plan of Thermostatic scenario}
	\label{fig:BSSResult}
\end{figure*}

Fig.~\ref{fig:PowerScheme-a} demonstrates the power scheme on the commercial side of LES, which is the main dispatch target with the majority of V2G potential. Through the temporal inconsistency of charging and discharging of ES and EVBs, the fixed load can be properly shifted according to the varying real-time price. In Fig.~\ref{fig:PowerScheme-a}, even though BSS requires large energy consumption for refueling swapped EVB, it still concentrates power import in the valley price period, indicating the marginal benefits of V2G load shifting arbitrage exceeds the revenue from swapping business. Besides, excess solar irradiation is fully utilized due to the storing ability in LES. 

The LPC results estimate the dynamics and characterize the power performance of battery during the solution of dispatch. Taking the single EVB of $ j_g=1,g=1 $ as the example, it is revealed in Fig.~\ref{fig:EVResult-a} that the power is limited in a dynamic SOPT region (grey shaded). And the SOPT boundary is essentially a mapping of varying SOC according to the thermostatic SOPT estimation from Fig.~\ref{fig:SOPT_line}, which is marked in the SOC-SOPT coordinate.

For BSS dispatch, the EVB power in BSS is similarly limited by the dynamic SOPT as EV. However, as shown in Fig.~\ref{fig:BSSResult-a}, the advent of online or offline signal reverses BSS occupying state, and battery physical parameters are kept at zero value when unoccupied. SOPT of EVB in BSS is a discontinuous function, corresponding to the fact that measurements e.g. SOC, SOPT, I are primarily battery states but artificially defined towards CDs.

\subsubsection{Case II: Dynamic temperature scenario}
This section investigates the efficacy of EM consideration in dispatch in addressing dynamic temperature. The temperature variation of EVBs can be attributed to two main factors: the first is heat generation and dissipation that occurs during EVBs' operation, and the second is the external ambient temperature change. In the day-ahead stage, the former can be considered continuous, while the latter is discretely distributed on different dates.

To evaluate the first factor, the EVBs' V2G available power is more aggressively exploited under a fast charging scheme (\textasciitilde1C) by setting a relatively looser constraint of energy conversion efficiency ($\eta_{min} = 98\% $), with the sacrifice of more heat generation and degradation. 
Under such a scheme, the cell temperature variation $ \Delta \Theta $ due to the internal chemical reaction and external converting equipment is not negligible. Thus heat dynamics along with SOPT is applied to characterize the dynamic cell temperature.

\begin{figure*}[htb]
	\centering
	\begin{subfigure}[t]{0.32\linewidth}
		\centering
		\includegraphics[width = 1\textwidth]{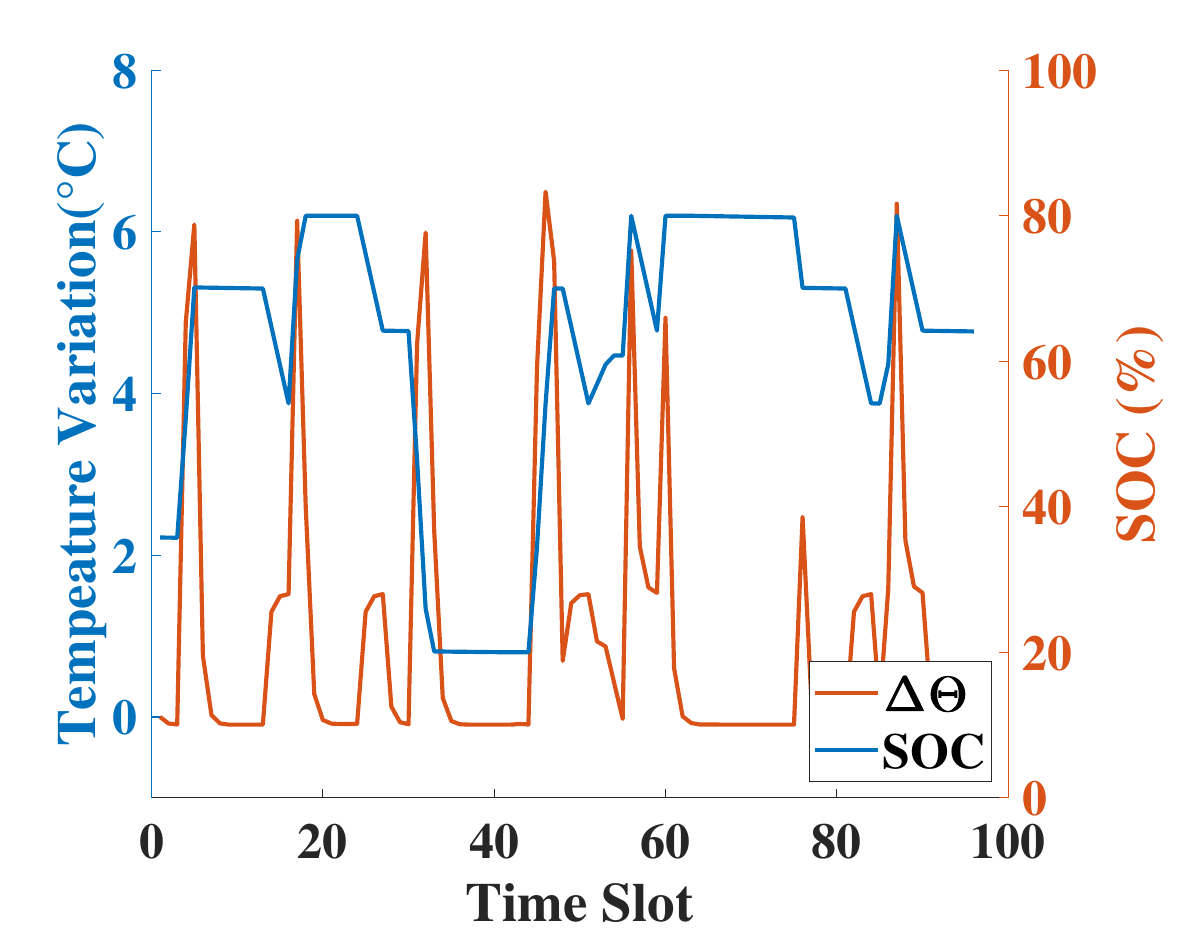}
		\subcaption{Cell Temperature Variation and SOC}
		\label{fig:EVResult_Temp}
	\end{subfigure}
	\begin{subfigure}[t]{0.32\linewidth}
		\centering
		\includegraphics[width = 1\textwidth]{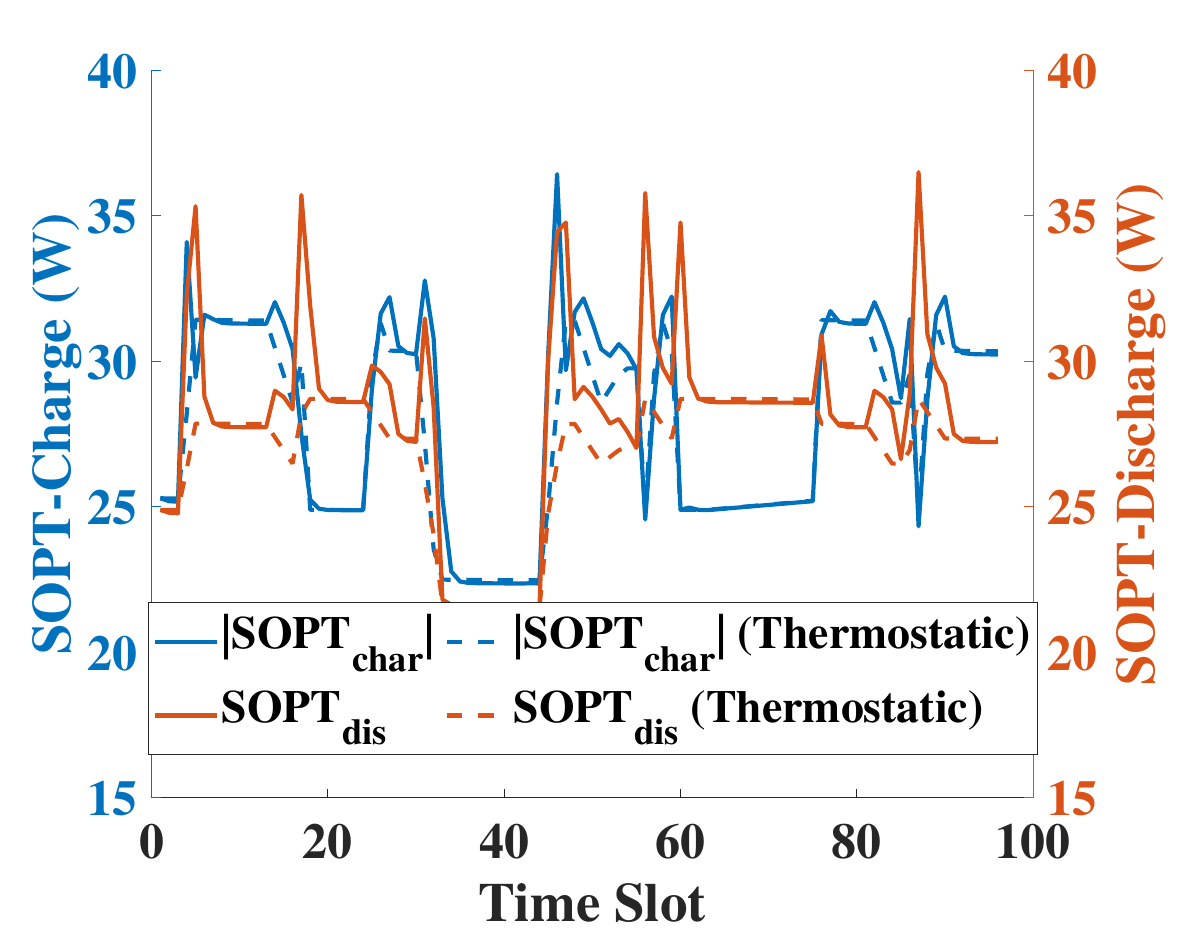}
		\subcaption{The Impact of Dynamic Cell Temperature}
		\label{fig:EVResult-SOPT}
	\end{subfigure}
	\begin{subfigure}[t]{0.32\linewidth}{
			\centering
			\includegraphics[width = 1\linewidth]{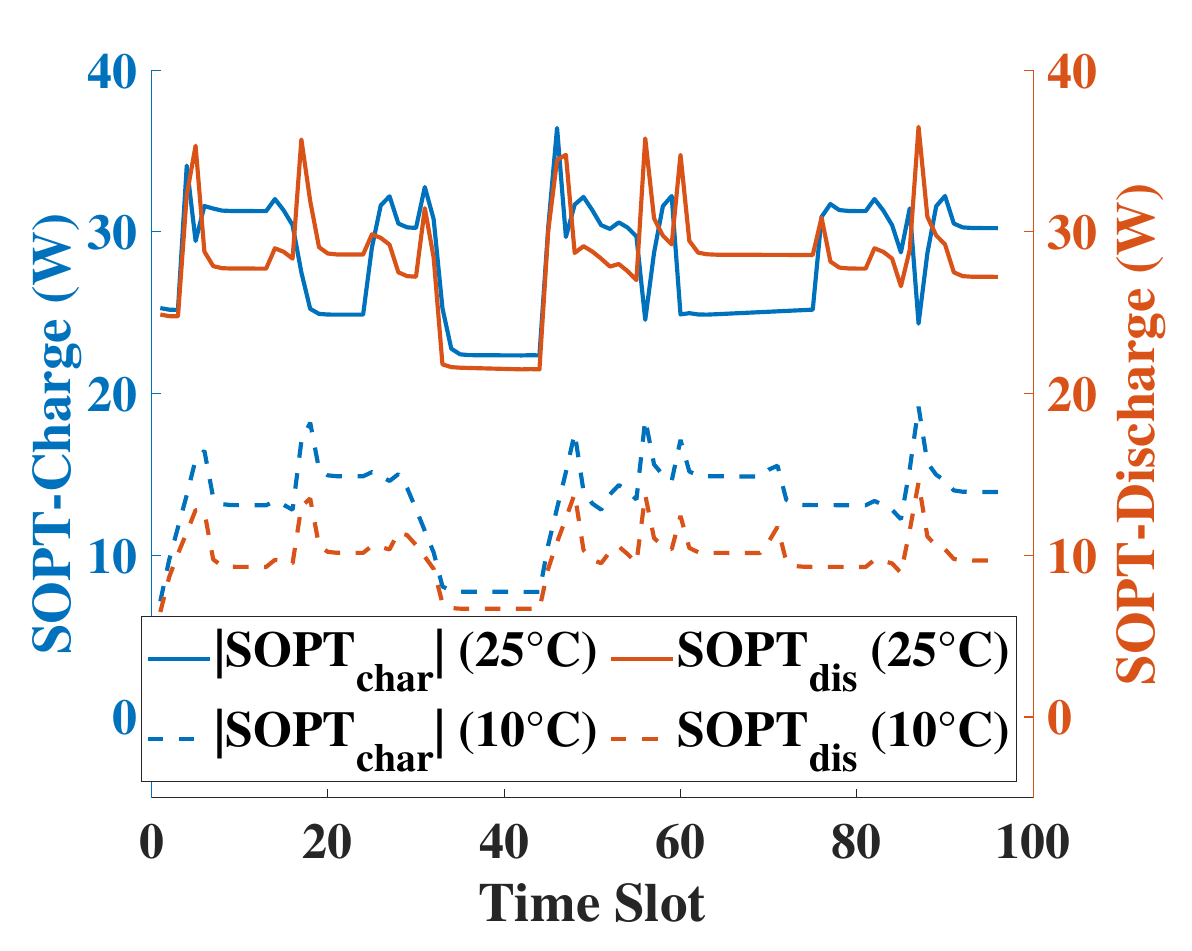}
			\subcaption{The Impact of $\Theta_{\amb}$}\vspace{6pt}
			\label{fig:EVResult_AmbTemp}}
	\end{subfigure}
	\centering
	\caption{EVB Dispatch under the Dynamic Temperature Scheme}
	\label{fig:EVResult_Case2}
\end{figure*}

Fig.~\ref{fig:EVResult_Temp} exhibits the SOC variation and the variation of temperature above ambient temperature. Since SOC varies as EVB power according to power dynamics, the temperature variation is related to the first-order derivative of SOC.

On the grid side, taking the dynamic temperature into account essentially affects the available power boundaries of EVBs, whose responses to temperature are modeled by the SOPT shown in Fig.~\ref{fig:SOPT}. According to \ref{sec:SOP}, SOPT is obtained under the combined influence of cell SOC and cell temperature. Referring to the SOC in Fig.~\ref{fig:EVResult_Temp}, the fundamental variation of SOPT in Fig.~\ref{fig:EVResult-SOPT} originates from the varying SOC, while the difference between the true SOPT and a comparative pseudo-thermostatic SOPT stems from the dynamic temperature.

For the second factor, ambient temperature $ \Theta_{\amb} $ serves as the initial setting of EVBs' cell temperature as well as the object of heat exchange during operating. Taking $ \Theta_{\amb} = 10^{\circ}C $ to represent a typical winter in south China, Fig.~\ref{fig:EVResult_AmbTemp} exhibits the SOPT comparison under different ambient temperatures when taking the first factor into account as well. 

Generally, EVBs will have a greater available power to be dispatched in the moderate environment than in a colder environment. But it should be noted that this phenomenon is not rigid since cells will be heated when being actively dispatched under low ambient temperature, which essentially offsets the effects of coldness.

\subsubsection{Case III: The effect of EVB aging}
To consolidate the importance of considering EM in dispatch, the efficacy of dispatching degraded cells is evaluated. EVB cycle-life profile indicates a descending battery performance \cite{hu_battery_2020}. Thus the degraded EVBs should be dispatched more cautiously to prevent accelerated degradation or even safety hazards. In LPC, the optimization-based SOPT is able to restrain such concerns implemented by the aging information from EM. By utilizing such SOPT results, the dispatch plan will respond to the varying cycle life of EVBs to reduce excessive use.

\begin{figure}[htb]
	\centering
	\includegraphics[width = 0.3\textwidth]{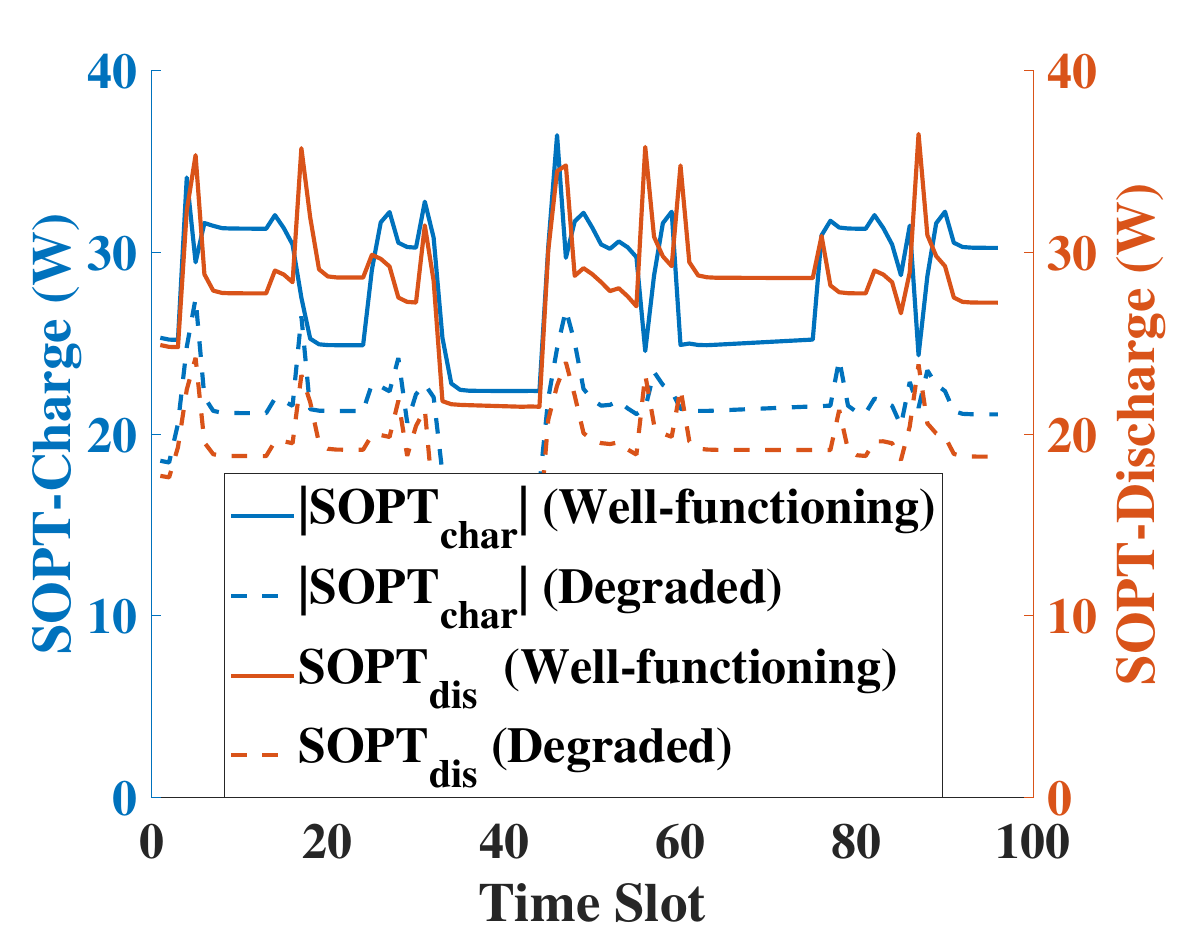}
	\caption{The EVB aging effect on SOPT}
	\label{fig:Cyclelife}
\end{figure}

A well-functioning EVB after 500 normal cycles and a degraded EVB after 3000 normal cycles are compared under the same case setting to validate the effect of aging on dispatch. Considering the aging mechanism, degraded EVB contains less active materials and lithium inventory, which will influence the available power to be dispatched regarding the grid's interest. Thus charging and discharging SOPT are compared in Fig.~\ref{fig:Cyclelife}. And it can be seen that by utilizing the EM, available power boundaries will descend as EVBs' aging, leading to a more conservative dispatching plan during V2G as the aging concern addressed by degraded EVBs.

\subsection{Model Assessment: Dispatch Model Evaluation}
\subsubsection{Peak shaving performance}
Peak shaving volume serves as an important indicator of the grid-connected EVs participating system level dispatch. Total discharged energy ($\sum_t P^t_{d,\LC}\Delta t$) from EVB is employed to estimate the peak shaving volume. Under $\Theta_{\amb}=10^{\circ}$C, the peak shaving volume of BSS significantly decreased by 65.48\% compared to due to the extra time it takes to charge EVB for retaining swapping demand with descended LiB SOPT. Charging EVBs are less affected by coldness with a slight 1.25\% drop of peak shaving volume, which may attribute to the fact that its dispatch plan possesses considerable non-operating moments and therefore the heat generation potential to increase the temperature for higher SOPT. Despite the reduced peak shaving volume, the total energy cost only inflates 2.45\% through the precise dispatch with EM consideration.

\subsubsection{Dynamics incorporation efficiency}
To evaluate the efficiency of implementing LPC with matrix-based state update methods in energy dispatch, the thermostatic case M2* is compared with dynamic temperature case M2 as a validation. As given in Table~\ref{tab:DATS-Summary}, although there exist more constraints and variables in M2 from the EM dynamics and their corresponding recursive update, the total optimization time of the two cases is within the same range and is approximately similar. The extra time taken by M2 seems consistent and is mainly cost by matrix formulation and model preprocessing.

\subsubsection{Dispatch model efficiency}
To validate the importance of properly formulating the dispatch problem with BSS integer logic when considering EM and to assess the efficiency of M2, a quasi-comparison study of M2 against M1 is conducted since there have been no comparable models in the literature. As given in Appendix.~\ref{apx:BSS-M1}, M1 applies the dispatch model with a widely-used battery swapping logic from \cite{sarker_optimal_2015}, and it is integrated directly with the recursive constraints with no specific design. Furthermore, to investigate the source of increasing efficiency, M1* is the contrast model that adopts similar chronological aggregation with time stamp $ \Delta h $, whereas individual incorporation is not applicable in M1* since it cannot buffer the quantity difference between swapping demand and aggregated logic as the virtual warehouse in M2.

It is revealed in Table~\ref{tab:DATS-Summary} that M2 substantially outperforms the other two M1 in terms of elapsed time. In the 500 EVBs case, M2 saves 72.63\% solving time compared to M1, while also exhibiting the capability to handle even larger scale problems than M1 as well. The results validate the model feasibility of practical application with a rather acceptable computing time in the day-ahead stage regarding the benefits of enhanced EVB precise control. The case study observation above coheres original intention of M2 and the remark on model design in Section~\ref{sec:Analysis}.

In addition, M1* exhibits a better performance than M1 in the large-scale problem, which proves the effectiveness of chronological aggregation.
And it should be noted that M1 has advantages after all because each EVB is independently dispatched on both logic and power, resulting in an exceeded ability to carry more swapping demand compared to M2.

\begin{table}[htb]
	\centering
	\caption{\centering The Solution Time of Dispatch Model (unit:s)}
	\begin{threeparttable}
		\begin{tabular}{ccccc}
			\hline
			\textbf{EVB Quantity in BSS}  & \textbf{100} & \textbf{250} & \textbf{500} & \textbf{750} \\ \hline
			\textbf{M2}  & 84.07  & 121.99 & 261.54 & \textbf{576.02}		\\
			\textbf{M2*} & 39.29  & 42.92  & 196.09 & \textbf{496.08}		\\ 
			\textbf{M1}  & 106.26 & 288.16 & 955.56 & \textbf{NA\tnote{**}} \\ 
			\textbf{M1*} & 101.02 & 248.38 & 743.84 & \textbf{1775.86} 		\\ \hline
		\end{tabular}
	\end{threeparttable}
	\label{tab:DATS-Summary}
	\begin{tablenotes}
		\footnotesize
		\item{M1*}: Chronologically aggregated; 
		\item{M2*}: Thermostatic;
		\item{NA**}: Unable to reach gap limit after 3600s
		\item MILP Gap Limit = 5\%
	\end{tablenotes}
\end{table}

\subsection{Model Assessment: Battery Model Comparison}
To assess the necessity and superiority of considering EM in dispatch, a comparative study on the battery models is conducted under the fast charging scheme with dynamic cell temperature. To validate the efficacy of EM, typical LiB models including the SSM and ECM are compared. For ECM, a similar thermostatic SOPT estimation method is applied on a second-order RC circuit to obtain available power.

\begin{figure}[htb] 
	\centering
	\begin{subfigure}[t]{0.67\linewidth}{
			\centering
			\includegraphics[width = 1\linewidth]{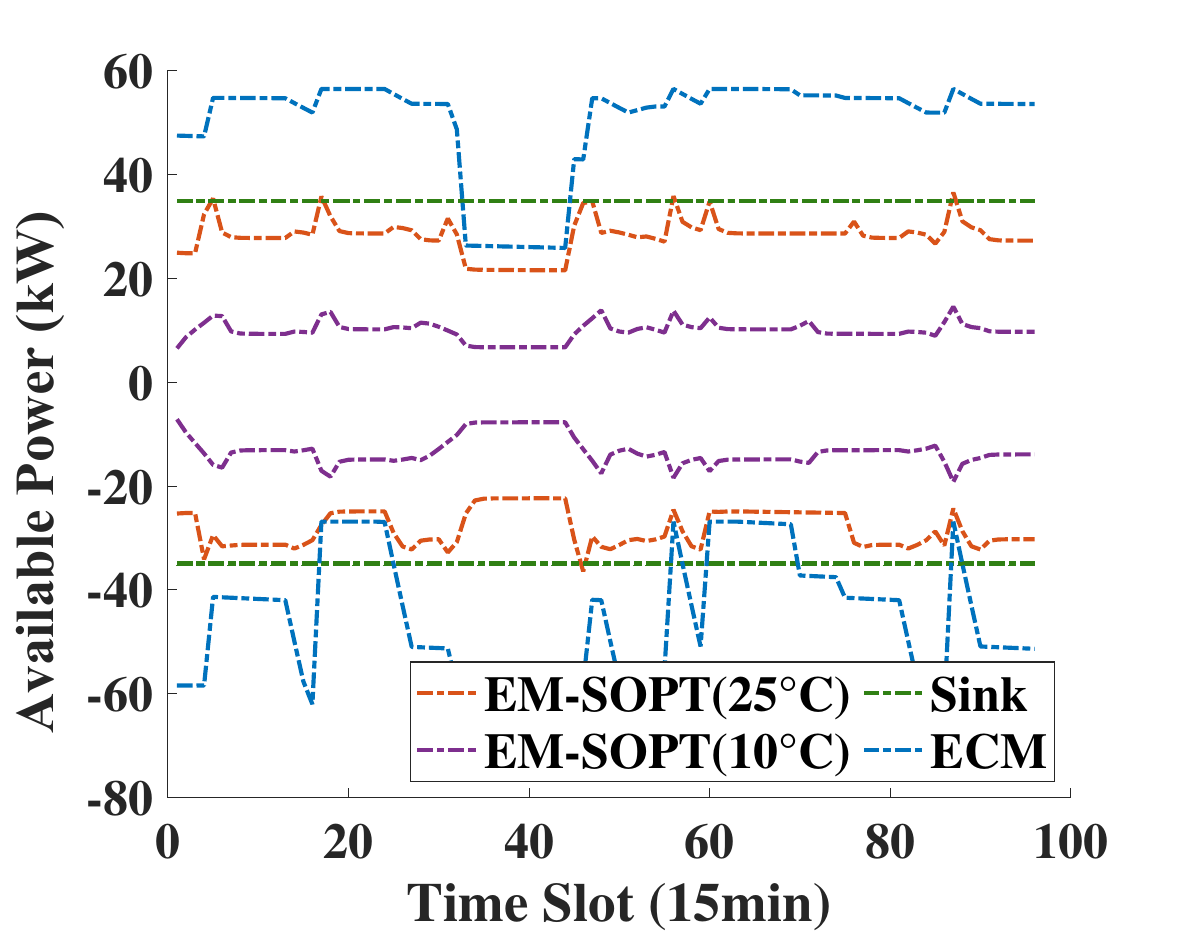}
			\subcaption{EVB Available Power}\vspace{6pt}
			\label{fig:LiBModelComparison-a}}
	\end{subfigure}
	\begin{subfigure}[t]{0.67\linewidth}{
			\centering
			\includegraphics[width = 1\linewidth]{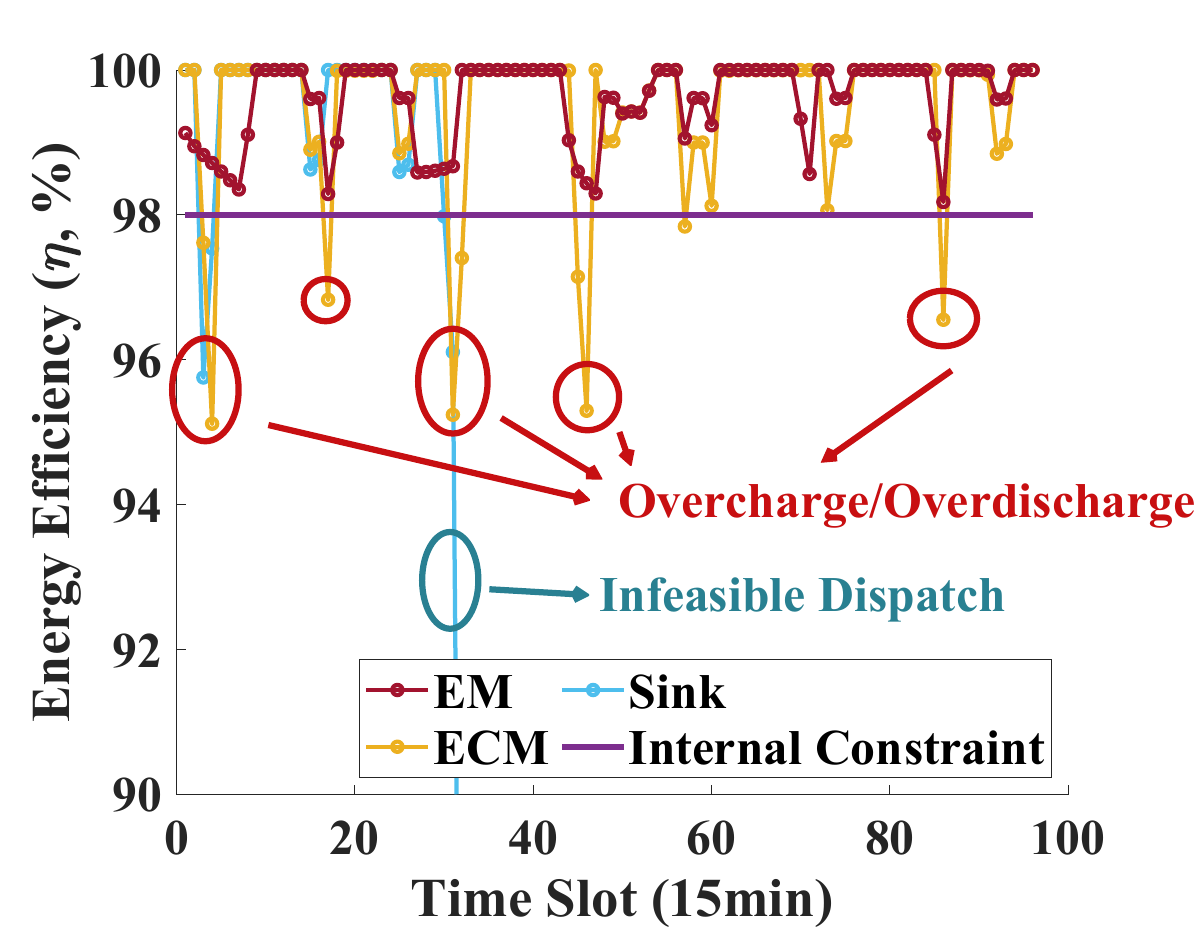}
			\subcaption{EVB $ \eta $}\vspace{6pt}
			\label{fig:LiBModelComparison-b}}
	\end{subfigure}
	\centering
	\caption{Battery Model Comparison on LiB Performance}
	\label{fig:LiBModelComparison}
\end{figure}

The available V2G power of the single EVB is compared in Fig.~\ref{fig:LiBModelComparison-a}. SSM limits power in a fixed boundary empirically set by designers, leading to the paradox to be over excessive or conservative. ECM can only partly identify the dynamics in available power, and the accuracy is inadequate due to the lower simulating precision and the omission of reaction-related constraints and dynamics. Whereas SOPT from EM is adaptive towards varying battery states as well as the operating conditions, which can spontaneously provide a much tighter SOPT limitation under coldness for instance.

To further evaluate the impact of employing different models, we reevaluate the entire day-ahead dispatch by full-order EM simulation with the obtained dispatch plans when considering different battery models. And to elaborate on the evaluation, we compare different battery models under multiple operating conditions as in Table.~\ref{tab:LiBModel-Summary}. The comparative study validates from three aspects:

\begin{enumerate*}[itemjoin=\\\hspace*{\parindent}]
	\item Dispatch feasibility. The results show that the dispatch plan when considering SSM under the fast charge scheme is physically infeasible indeed, because the SSM dispatch is not aware of draining out of EVB energy. It designates to continue discharging when the lithium concentration reaches the lower bound, since the current calculation and SOC update by SSM lack accuracy. On the contrary, EM consistently provides a feasible dispatch plan.
	
	\item Efficiency and heat. Cell energy efficiency $\eta$ is an important indicator for evaluating the long-term battery operation and the stability of participation in V2G service. And it explains most of the heat generation and reflects the operating status. Fig.~\ref{fig:LiBModelComparison-b} gives an example of the validation on EVB $\eta$, in which the $\eta$ is limited above 98\% effectively as the set restriction during LPC of fast charging scheme. But it can be seen that applying either SSM or ECM occasionally exceeds the expected limitation and operates in inefficient regions. Such inefficiency suggests that LiB endures overcharge or over-discharge constantly, leading to accelerated degradation or even thermal runaway. As for the adaptability under different operating conditions as suggested in Table.~\ref{tab:LiBModel-Summary}, considering EM in the dispatch consistently preserves high conversion energy efficiency, limiting $\eta_{\min}$ above the operating restriction $\boldsymbol{\Omega}$ in LPC adaptively. Other than the improved battery performance, the superiority of high efficiency can be more intuitively concluded by comparing cell internal heat generation $H_i$. For instance, when considering EM in dispatch, $H_i$ is significantly reduced by 69\% compared to utilizing ECM in the degraded cell case, which is attributed to raising the $\eta_{\min}$ from 95.12\% to 98.18\%.
	
	\item Aging rate. As it can be seen from Table.~\ref{tab:LiBModel-Summary}, the dispatch considering EM will be able to suppress excess aging proactively and adaptively, which is incapable in the dispatch utilizing ECM or SSM. The accumulative EVB capacity loss $C_L$ is employed to validate the capability of suppressing aging development.  the implementation of EM demonstrates significant efficacy in mitigating EVB degradation in energy dispatch, particularly for low-temperature conditions and degraded batteries. In such scenarios, applying EM achieves a reduction of 14.41\% and 24.19\% in terms of accumulative capacity loss compared to ECM, respectively.
\end{enumerate*}

Thus, applying EM is necessary and superior as the following considerations:
\begin{itemize}
	\item The feasibility of dispatch plans with EM consideration is adequate as it will not overestimate the remaining EVB energy. Consequently, it will not result in an inflated approximation of the energy cost reduction, and is practical to implement the dispatch on real facilities.
	\item ECM and SSM cannot provide information about the inside chemical reaction. This makes it less bounded and generates a radical available V2G power, which may contribute to excess aging and less cell efficiency with associated extra heat or even safety hazards.
	\item SSM is a purely static model and ECM is statically fitted, so neither is able to modify the dispatch adaptively under different states, ambient temperatures, anode materials, and degradation levels. While considering EM is sufficient to model the actual dynamic performance adaptively under real operating conditions.
\end{itemize}

\begin{table}[htb]
	\caption{Summary of Battery Model Comparison}
	\begin{tabular}{lllll}
		\hline
		\textbf{}                                       & \textbf{}         & \textbf{SSM} & \textbf{ECM} & \textbf{EM}     \\ \hline
		\multirow{3}{*}{\textbf{Normal Temperature}}    
		& $\eta_{\min}$(\%) & NA           & 97.64        & \textbf{98.64}  \\
		& $H_i$(kJ) 		& NA           & 0.7483       & \textbf{0.5774}	\\
		& $C_L$(mAh)        & NA           & 1.0666       & \textbf{0.9743} \\ \hline
		\multirow{3}{*}{\textbf{Low Temperature}}       
		& $\eta_{\min}$(\%) & NA           & 97.4         & \textbf{98.48}  \\
		& $H_i$(kJ) 		& NA           & 0.8303       & \textbf{0.6465}  \\
		& $C_L$(mAh)        & NA           & 0.5973       & \textbf{0.5112} \\ \hline
		\multirow{3}{*}{\textbf{Well-Functioning Cell}} 
		& $\eta_{\min}$(\%) & NA           & 96.67        & \textbf{98.45}  \\
		& $H_i$(kJ) 		& NA           & 1.1458       & \textbf{0.6905}  \\
		& $C_L$(mAh)        & NA           & 0.2983       & \textbf{0.2580}  \\ \hline
		\multirow{3}{*}{\textbf{Degraded Cell}}         
		& $\eta_{\min}$(\%) & NA           & 95.12        & \textbf{98.18}  \\
		& $H_i$(kJ) 		& NA           & 1.7265       & \textbf{0.5459}  \\
		& $C_L$(mAh)        & NA           & 0.3427       & \textbf{0.2598} \\ \hline
	\end{tabular}
	\begin{tablenotes}
		\footnotesize
		\item{NA}: Physically infeasible dispatch
	\end{tablenotes}
	\label{tab:LiBModel-Summary}
\end{table}

\section{Conclusion}
\label{sec:conclusion}
Grid-connected EVs are a viable solution for grid regulation, but the dynamic performance related to reaction mechanism of LiB makes the precise dispatch necessary. This article investigates the consideration of EM in the energy dispatch of grid-connected EVs, implementing the LPC to consider dynamic voltage, temperature, and available power. The state update is efficient and non-iterative in optimization with the matrix-based method. Referring to the recursive constraints given by LPC, CS and BSS are modeled accordingly under the consideration of optimization complexity. And the LES with high grid-connected EV penetration is investigated as the scenario to validate the efficacy of the proposed method.

The results reveal that the consideration of EM is adequate and the integration of EVs in the grid plays a pivotal role in reducing energy costs and PV curtailments. By utilizing EM, the dispatch is able to set state-dependent power limitations and is adaptive to cell temperature and degradation level of EVB. In comparison to ordinary model, the proposed dispatch model M2 not only exhibits a substantial 72.63\% reduction in terms of the computational time when integrating LPC results, but also outperforms in handling large-scale problems. Besides, battery model comparison reveals the necessity and superiority of considering EM in terms of dispatch feasibility and maintaining the anticipated EVB energy conversion efficiency under various conditions to reduce internal heat generation. Considering EM in dispatch also inhibits aging and reduces capacity loss up to 24.19\%. The assessment of the dispatch model demonstrates that the proposed model is sufficient for application in day-ahead dispatch, indicating its suitability for practical implementation.

Our future work includes long-term planning of energy storage and EVB life cycle value assessment considering a high-precision battery model.
\ifCLASSOPTIONcaptionsoff
\newpage
\fi

\appendix
\subsection{Intuition-oriented BSS Optimization Model (M1)}
\label{apx:BSS-M1}
To validate the M2's efficiency with EM consideration, an ordinary model M1, which is intuition-oriented, is compared against. M1 encounters the same complexity of considering EM, which adopts a widely-used battery swapping logic from \cite{sarker_optimal_2015} integrated directly with the recursive constraints from LPC with no specific design. In this appendix, the logic of BSS is briefly presented as follows.

Different from M2, M1 designates the battery swapping pair one by one. And $k$ in M1 represents the index of EVB instead. Denote the swapping signal of each in-station EVB by $ w^{k,t} $, where $ w^{k,t}=0 $ represents that $ k $-th in-station full-charged EVB are swapped by an out-station depleted EVB at $t$, while $ w^{k,t}=1 $ represents that EVB remains in-station. At $t$, the required $ Q_{\req}^t $ should be consistent with the total quantity of swapping pairs ($ w^{k,t} = 0 $):
\begin{equation}
	\label{eqa:M1-Q_Req}
	\sum_{k=1}^{K}1-w^{k,t} = Q_{\req }^t
\end{equation}

For each swapping pair: depleted EVBs are drained of energy, assuming the SOC to be $ SOC_{\BSS,E} $ at the moment of swapping, while full-charged EVBs need to meet the SOC requirement $ SOC_{\BSS,F} $ for external swapping services at the previous moment of swapping: 
\begin{equation}
	y^{k,t} = w^{k,t}(1-w^{k,t+1}) ,~\forall t \leq T-1
\end{equation}
\begin{equation}
	y^{k,t}SOC_{\BSS,F} \leq SOC_{\BSS}^{k,t}
	\label{eqa:M1-SOC_Req}
\end{equation}
where $ y^{k,t} $ denotes the swapping signal turning 0 from 1.

Then the final optimization of M1 regarding the LES dispatch is:
\begin{equation}
	\begin{aligned}
		&\begin{aligned}
			\text{(M1)~minimize:}   \quad Z &= Z_{\CST,C}+Z_{\CST,R}-Z_{\APR,\ES} \\&-Z_{\APR,\EV}-Z_{\APR,\BSS}-Z_{\REV} \\
		\end{aligned} \\
		&\begin{aligned}
			\text{subject to:}~(\ref{eqa:SOC-I_0}),(\ref{eqa:SOC-Update-cell}),(\ref{eqa:Temp-Update-cell}),(\ref{eqa:SOP-Est}), (\ref{eqa:EV-I})-(\ref{eqa:EV-Unplug}), (\ref{eqa:Grid_C})-(\ref{eqa:M1-SOC_Req})\\
		\end{aligned}
	\end{aligned}
\end{equation}

\bibliographystyle{IEEEtran}
\bibliography{ref}

\begin{IEEEbiography}[{\includegraphics[width=1in,height=1.25in,clip,keepaspectratio]{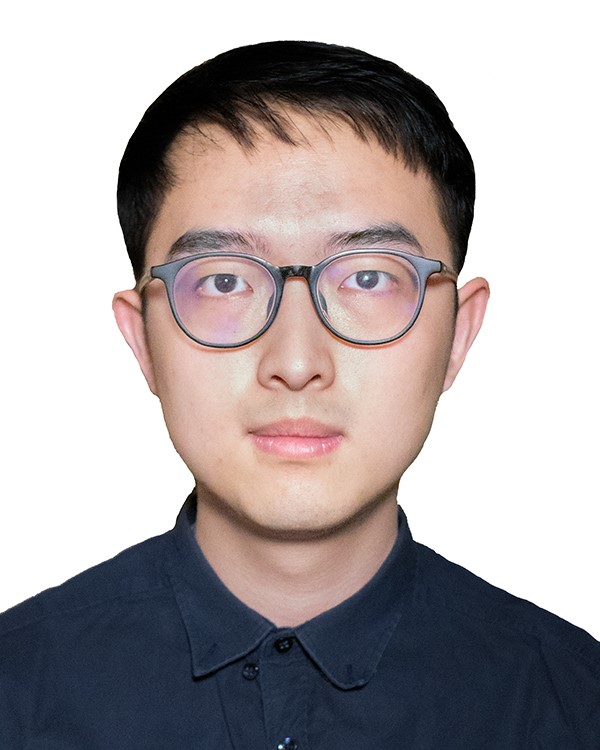}}]{Yuanbo Chen (Student Member, IEEE) received a B.S. degree in electrical engineering from Tsinghua University, Beijing, China, in 2022, where he is currently pursuing a Ph.D. degree. His research interests include lithium-ion battery management, battery charging and swapping optimization.}
\end{IEEEbiography}

\begin{IEEEbiography}[{\includegraphics[width=1in,height=1.25in,clip,keepaspectratio]{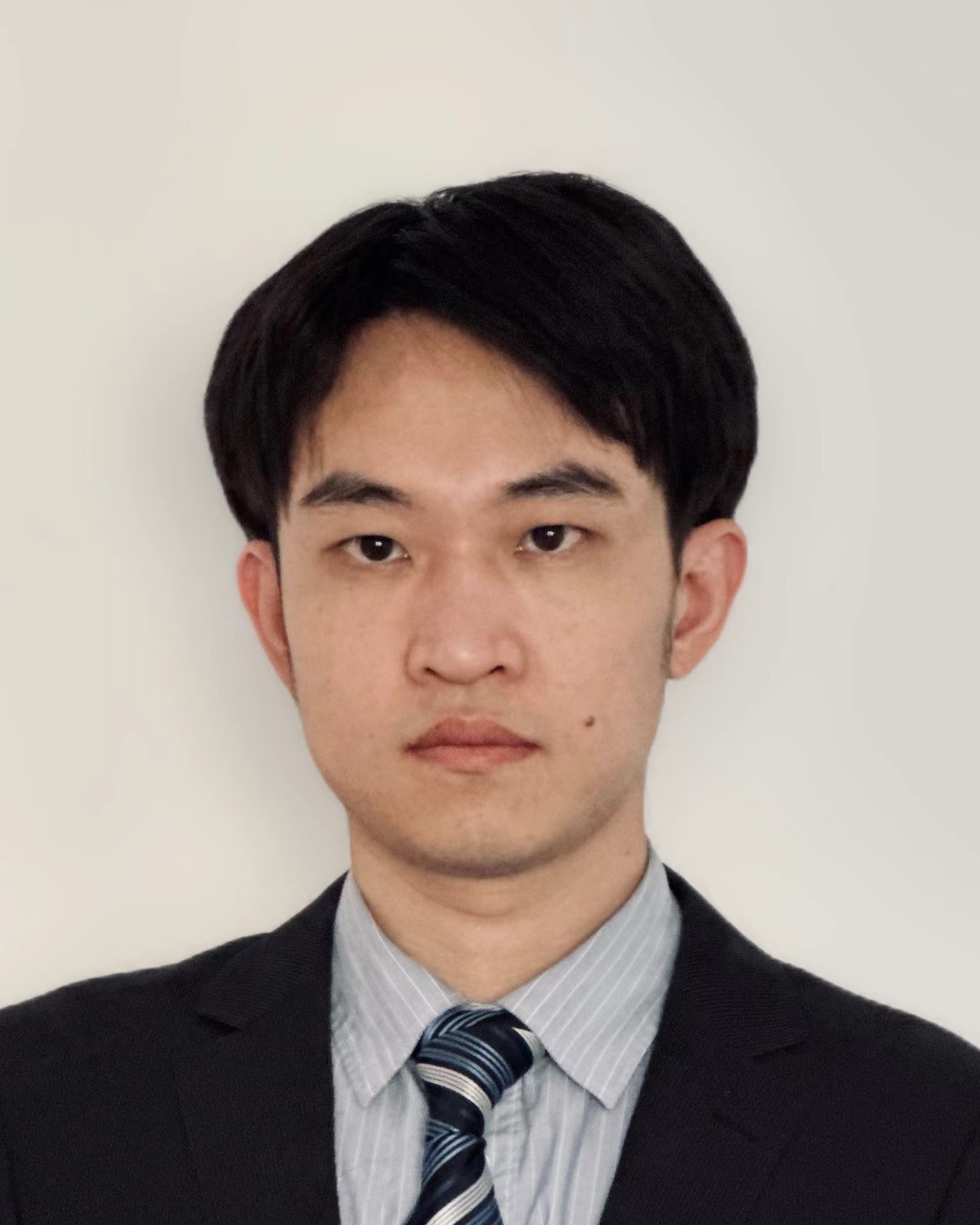}}]{Kedi Zheng (Member, IEEE) received the B.S. and Ph.D. degrees in electrical engineering from Tsinghua University, Beijing, China, in 2017 and 2022, respectively. He is currently a Postdoctoral Researcher with Tsinghua University. He is also a Visiting Research Associate with The University of Hong Kong. His research interests include data analytics in power systems and electricity markets.}
\end{IEEEbiography}

\begin{IEEEbiography}[{\includegraphics[width=1in,height=1.25in,clip,keepaspectratio]{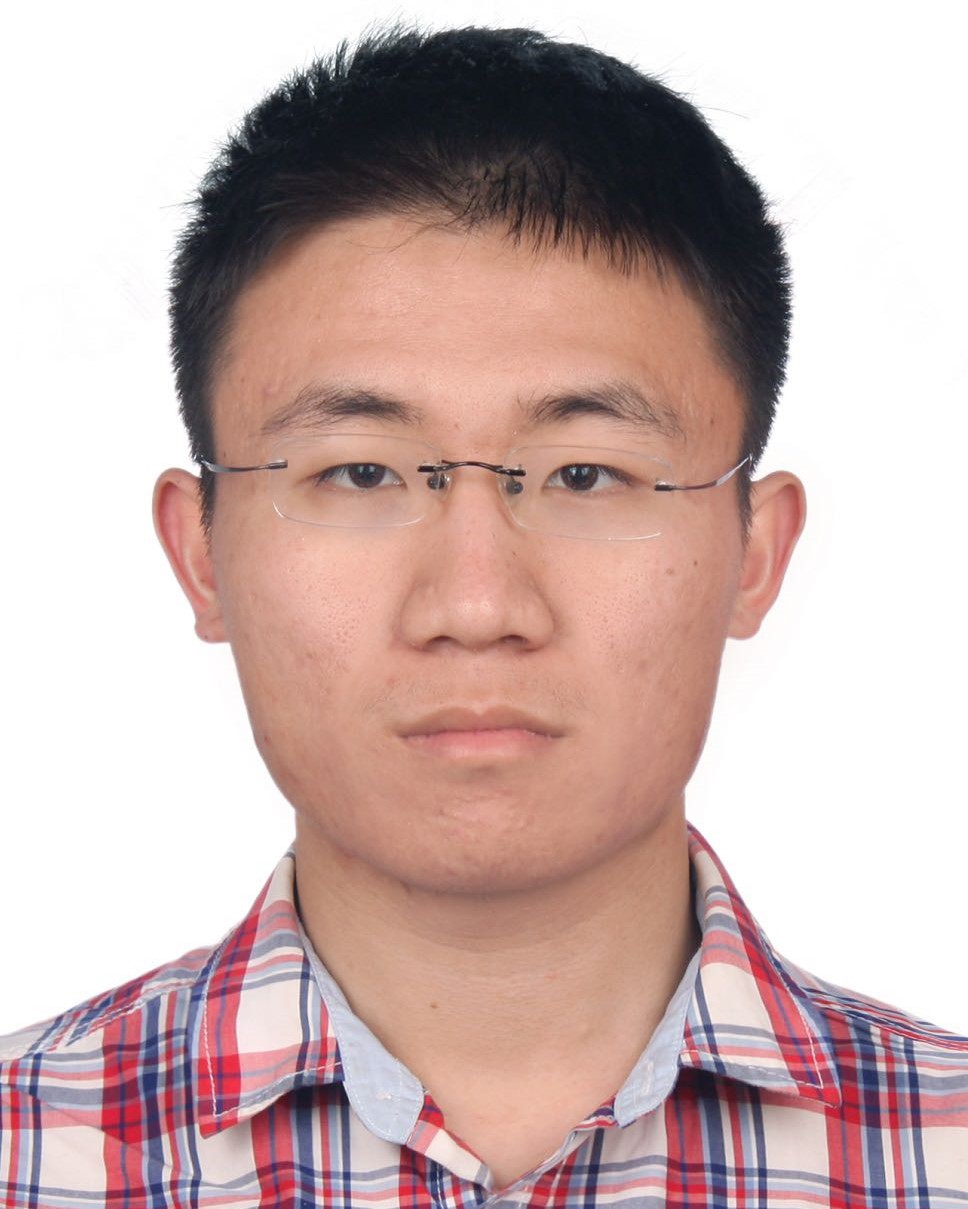}}]{Yuxuan Gu received a B.S. and Ph.D. degrees in electrical engineering from Tsinghua University, Beijing, China, in 2018 and 2023, respectively. His research interests include lithium-ion battery modeling and management, big data technology, deep learning, the electricity market, and optimization.}
\end{IEEEbiography}

\begin{IEEEbiography}[{\includegraphics[width=1in,height=1.25in,clip,keepaspectratio]{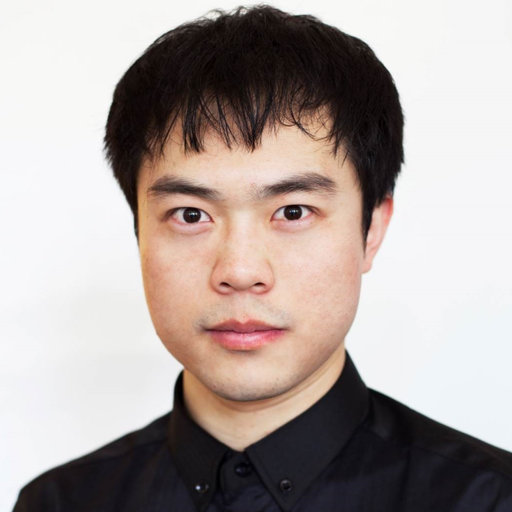}}]{Jianxiao Wang (Member, IEEE) received the B.S. and Ph.D. degrees in electrical engineering from Tsinghua University, Beijing, China, in 2014 and 2019, respectively. He is currently an Assistant Researcher with National Engineering Laboratory for Big Data Analysis and Applications, Peking University, Beijing, China. His research interests include smart grid operation and planning, hydrogen and storage technology, transportation and energy systems integration, electricity market, and data analytics.}
\end{IEEEbiography}

\begin{IEEEbiography}[{\includegraphics[width=1in,height=1.25in,clip,keepaspectratio]{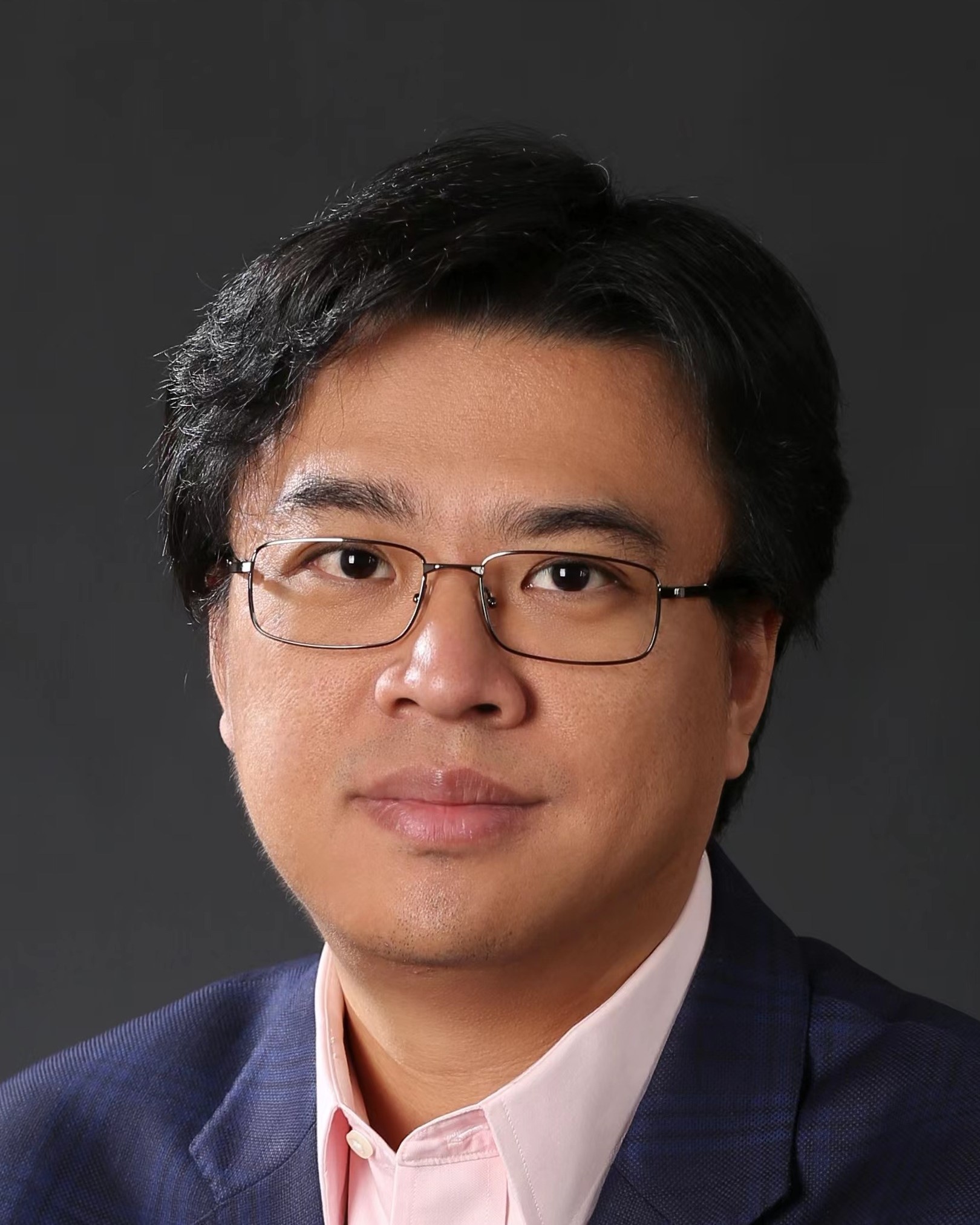}}]{Qixin Chen (Senior Member, IEEE) received a Ph.D. degree from the Department of Electrical Engineering at Tsinghua University, Beijing, China, in 2010. He is currently a Professor at Tsinghua University. His research interests include electricity markets, power system economics and optimization, low-carbon electricity and power generation expansion planning.}
\end{IEEEbiography}
\vfill
\end{document}